\documentclass[prd,preprintnumbers,showkeys, nofootinbib]{revtex4}
\usepackage{graphicx,amsmath,amssymb,epsfig,slashed}
\newcommand{\mbf}[1]{\mbox{\boldmath $#1$}}

\renewcommand{\Im}{\operatorname{Im}}
\newcommand{\tr}{\operatorname{tr}}
\newcommand{\Tr}{\operatorname{Tr}}
\newcommand{\cL}{{\cal L}}
\newcommand{\cN}{{\cal N}}
\newcommand{\cO}{{\cal O}}
\newcommand{\bk}{\mbf{k}}
\newcommand{\bq}{\mbf{q}}
\newcommand{\bl}{\mbf{l}}

\newcommand{\beps}{\mbf{\epsilon}}
\newcommand{\bN}{\mbf{N}}
\newcommand{\lambdab}{\bar{\lambda}}
\newcommand{\SUN}{$SU(N_c)$}
\newcommand{\SUR}{$SU_R(4)$}

\begin{document}

\preprint{DESY 08-044}

\title{Four point function of $R$-currents in ${\cal N}=4$ SYM \\in the Regge limit at weak coupling} 

\author{J.~Bartels}
\email{bartels@mail.desy.de}
\author{A.-M.~Mischler}
\email{anna-maria.mischler@desy.de}
\author{M.~Salvadore}
\email{michele.salvadore@gmail.com}

\affiliation{I\,I. Institut f\"ur Theoretische Physik, Universit\"at Hamburg,\\ 
Luruper Chaussee 149, 22761 Hamburg, Germany}


\begin{abstract}
 We compute, in $\cN=4$ super Yang-Mills theory, the four point correlation function of  $R$-currents in the Regge limit in the leading logarithmic approximation at weak coupling. Such a correlator is the closest analog 
to photon-photon scattering within QCD, and there is a well-defined procedure
to perform the analogous computation at strong coupling via the AdS/CFT correspondence.
The main result of this paper is, on the gauge theory side, 
the proof of Regge factorization and the explicit computation of the
$R$-current impact factors.
\end{abstract}
\maketitle

\section{Introduction}

There are many aspects of QCD that are still lacking a satisfying
understanding from first principles.
One is the behavior in the Regge limit, where the theory is expected
to be better formulated in terms of new effective fields, the Reggeized
particles \cite{Gribov:1968fc,Lipatov:1996ts}. One of the central building blocks 
of this Reggeon field theory is the Balitsky-Fadin-Kuraev-Lipatov (BFKL) Pomeron which comes as a bound state of two 
Reggeized gluons with vacuum quantum
numbers \cite{Kuraev:1976ge-Kuraev:1977fs-Balitsky:1978ic}.
While the original calculations were done in the leading 
logarithmic approximation (LLA), the requirement of
high precision has led to the computation of subleading corrections (NLO corrections) 
to the BFKL equation \cite{Fadin:1998py,Camici:1997ij-Ciafaloni:1998gs},
and they have been found to be large. While, for finite values of $N_c$
further steps beyond NLO will extend beyond the ladder structure and hence open 
the full complexity of Reggeon field theory, there is evidence that 
the large-$N_c$ limit suppresses the transition from two to four 
Reggeized gluons and thus allows, also beyond the NLO corrections, to stay 
within the ladder approximation.        

Beside its phenomenological relevance, high energy physics
has been a prolific source of theoretical cues.
In the early days, the proposal by Veneziano
\cite{Veneziano:1968yb} of crossing-symmetric,
Regge behaved amplitude turned out to be a key
point for the beginning of the string theory era.
Later on, in the early nineties, when studying unitarity
corrections to the BFKL Pomeron, Lipatov \cite{Lipatov:1993yb,LiFK} found
the first occurrence of integrable structures in
four dimensional quantum field theories: In the large-$N_c$ limit, the generalization of the 
BFKL evolution equation, the Bartels-Kwiecinski-Praszalowicz (BKP) evolution equations \cite{BKP} for the $n$ gluon 
state, were found to be integrable. 

Recently, the connection between quantum field theory 
and string theory was revived by the advent of the AdS/CFT correspondence
\cite{Maldacena:1997re-Witten:1998qj-Gubser:1998bc}.
This conjectured connection between Yang-Mills theories
(the maximally supersymmetric version of QCD,
$\cN=4$ super Yang-Mills theory (SYM), at large $N_c$, being the most attractive example) 
and some string theory (type IIB on $AdS_5 \times S^5$ for the case
just mentioned) has motivated, among other directions of interest,
also the analysis of the high energy limit in supersymmetric theories,
in particular the BFKL Pomeron 
\cite{Kotikov:2000pm-Kotikov:2002ab-Kotikov:2003fb-Kotikov:2004er}
and the vacuum singularity \cite{fromJaniktoHatta}. 

On the gauge theory side, the most reliable environment of investigating
the Pomeron is provided by the scattering of electromagnetic currents,
e.g., the total cross section of the scattering of two virtual photons
\cite{Bartels:1996ke,Brodsky:1997sd}.  
In order to be able to define correlation functions that are defined 
on both the gauge theory and the string theory side, it has been 
suggested \cite{CaronHuot:2006te} to use, as a substitute of the electromagnetic current, 
the $R$-currents belonging to the global $SU_R(4)$ of the $\cN=4$ SYM theory. 
To be more precise, one picks a $U(1)$ subgroup of the $SU_R(4)$ group. 
It therefore seems natural to investigate four  point correlators 
(and even $n$ point correlators with $n>4$) of these $R$-current operators,
representing correlation functions which are well-defined 
both on the gauge theory and the string theory side.
Whereas two point and three point correlators of the $R$-current operators 
have been studied before \cite{Chalmers:1998xr}, an analysis of four point correlation functions 
has not yet been performed.

In this paper we address, within $\cN=4$ SYM, the Regge limit of $R$-current operators, 
beginning with the gauge theory side. In QCD it is well known that, in the high energy 
Regge limit, the four point amplitude of the electromagnetic current 
factorizes into impact factors of the (virtual) photon and the BFKL 
Green's function that describes the energy dependence.
In this paper, as a start, we will verify that this expectation remains valid 
also for the supersymmetric extension, where scalar fields have to be 
included, and the fermions belong to the adjoint representation of the 
gauge group. Since the $R$-currents are non-Abelian, their associated Ward identities are more complicated then in QED, and this causes some subtleties in the treatment of UV divergencies.     
We investigate the one-loop box diagrams and compute, in the leading 
logarithmic representation, the impact factors of the $R$-current\footnote{In a 
recent paper \cite{Cornalba:2008qf} the impact factors of scalar currents have been 
computed.}. 
Since, in the leading logarithmic approximation, the BFKL Green's function 
remains the same as in the nonsupersymmetric case, we thus find the 
supersymmetric analog of the $\gamma^* \gamma^*$ total cross section 
discussed in QCD. In a forthcoming  paper we will turn to the dual analog on the string 
theory side where the graviton is expected to play the dominant role. 
         
\section{Review of photon-photon scattering in QCD}

The most convenient way of addressing Regge dynamics in
QCD is the study of the elastic scattering of two highly virtual photons.
The large virtuality of the external photons provides hard scales
that allow us to use perturbation theory.
One focuses on the computation of the leading order in the electric
charge $\alpha$, at which each photon splits into a quark-antiquark
pair, but the order in the strong coupling constant $\alpha_s$
can be arbitrary high. The decay of the photon is mediated by the
electromagnetic current $j_\mu$ associated with the $U(1)$ gauge symmetry
of QED. Therefore the computation reduces to evaluating the four
point correlation function of this current.
In momentum space it reads\footnote{Note that $p_{A,B}$ are taken
to be incoming while $p_{A,B}'$ are outgoing.}
\begin{eqnarray}
  \label{eq:4pointfuncQCD}
    &&i (2\pi)^4 \delta^{(4)}(p_A+p_B-p_{A'}-p_{B'}) A(s,t)^{\mu_A \mu_B \mu_{A'}  \mu_{B'}} =\nonumber\\
    &&\qquad\int \prod_i d^4x_i~
    e^{- i p_A \cdot x_A - i p_B \cdot x_B
    +i p_{A'} \cdot x_{A'} + i p_{B'} \cdot x_{B'} }\nonumber\\
  &&\qquad\times\langle j^{\mu_A}(x_A) j^{\mu_B}(x_B)
    j^{\mu_{A'}}(x_A') j^{\mu_{B'}}(x_B') \rangle \, ,
\end{eqnarray}
where $A$ depends upon the external momenta only through the usual
Mandelstam variables\footnote{Bold symbols label
2-dimensional transverse vectors, $\bk=(k_1,k_2)$.}
\mbox{$s = (p_A + p_B)^2 > 0$}, $t=q^2=(p_A-p_A')^2 \simeq -\bq^2 < 0$,
and the virtualities of the current momenta $Q_i^2 = - p_i^2 > 0$.
The Regge limit is defined as
\begin{equation}
  \label{eq:ReggeLim}
  s >> |t|, Q_i^2 \, .
\end{equation}
We will perform the computation using the Sudakov decomposition of momenta
discussed in the appendix.
It is convenient to compute the amplitude \eqref{eq:4pointfuncQCD}
in terms of its projections onto the polarization vectors of the
external photons.
The reader is referred to the appendix \ref{app:sudakovAndPol} for the explicit
definition of the polarization vectors in the Regge limit.
Once they are defined, we can use their completeness \eqref{eq:polComp}
in order to decompose the correlation function \eqref{eq:4pointfuncQCD} as
\begin{eqnarray}
  \label{eq:A4decompQCD}
 &&A(s,t)_{\mu_A \mu_B \mu_{A'} \mu_{B'}} = \nonumber\\
&&\qquad\sum_{\lambda_i}  \epsilon^{\lambda_A}_{\mu_A}(p_A)^* \epsilon^{\lambda_B}_{\mu_B}(p_B)^*
  \epsilon^{\lambda_{A'}}_{\mu_{A'}}(p_{A'}) \epsilon^{\lambda_{B'}}_{\mu_{B'}}(p_{B'})\nonumber\\
 &&\qquad\times \langle \lambda_A \lambda_B | A | \lambda_{A'} \lambda_{B'} \rangle \, ,
  \quad \lambda_i=L,\pm \, ,
\end{eqnarray}
where
$\langle \lambda_A \lambda_B | A | \lambda_{A'} \lambda_{B'} \rangle$
are the projections of $A$ onto the appropriate polarization vectors.
In the following we will often suppress, for the scattering amplitude $A$ on the LHS, the 
tensor indices.  

\subsection{Ward identities}
\label{sec:WI}

Let us briefly recapitulate the derivation of the Ward identities for the
time-ordered product of a conserved current, $j^{\mu}$ (satisfying  
$\partial_\mu j^\mu(x)=0$), with some other operators $\cO_i$.
Because of the theta-functions inserted by the time-ordering operator $T$,
there are terms proportional to delta-functions of time differences,
\begin{eqnarray}
  \label{eq:WardId1}
 && \partial_\mu T j^\mu(x) \cO_1(x_1)...\cO_n(x_n)=\nonumber\\
 &&\sum_{i=1}^n \delta(x^0-x_i^0) T \cO_1(x_1)...[j^0(x),\cO_i(x_i)]...
  \cO_n(x_n) \, .
\end{eqnarray}
From the standard commutation relation one sees that the equal-time
commutator of the zero-component of the current with an operator is
proportional to the charge of the operator itself under the symmetry
group,
\begin{equation}
  \label{eq:commJ0O}
  [j^0(\vec{x},t),\cO(\vec{y},t)] = \delta^{(3)}(\vec{x}-\vec{y})
  q_{\cO} \cO(\vec{x},t) \, .
\end{equation}
Here $q_{\cO}$ is the charge of the operator $\cO$ in units of electric
charge $e$.
Using \eqref{eq:commJ0O} in \eqref{eq:WardId1} one obtains the explicit
expression of the contact terms:
\begin{eqnarray}
  \label{eq:WardId2}
  &&\partial_\mu T j^\mu(x) \cO_1(x_1)...\cO_n(x_n)=\nonumber\\
  &&\qquad\sum_{i=1}^n \delta^{(4)}(x-x_i) q_{\cO_i}
  T \cO_1(x_1)...\cO_n(x_n) \, .
\end{eqnarray}
One sees then that there are no contact terms with neutral operators.
In particular, since in an Abelian theory the current is neutral,
there are no contact terms in the $T$-products of currents,
\begin{equation}
  \label{eq:WardId3}
  \partial_\mu T j^\mu(x) j^{\mu_1}(x_1)...j^{\mu_n}(x_n)=0 \, .
\end{equation}
Going to momentum space and taking the vacuum expectation value one gets
the well-known equation
\begin{equation}
  \label{eq:WardId4}
  p_\mu \langle j^\mu(p) j^{\mu_1}(p_1)...j^{\mu_n}(p_n) \rangle = 0 \, .
\end{equation}
Going from \eqref{eq:WardId3} to \eqref{eq:WardId4} involves a subtlety.
The integrations in the coordinates implied by Fourier transformation
pick up contributions from the regions where two or more currents are
at the same point. In some cases the product of currents at the same point requires some care, scalar QED is a simple example 
(Sec.\ref{sec:ScalarQED}).

\subsection{Box diagrams}

\begin{figure}
\begin{eqnarray}
  \label{eq:BoxesF}
  &&A^{(0)}_B=\nonumber\\
  &&\begin{minipage}{2.2cm}
    \includegraphics[width=2.2cm]{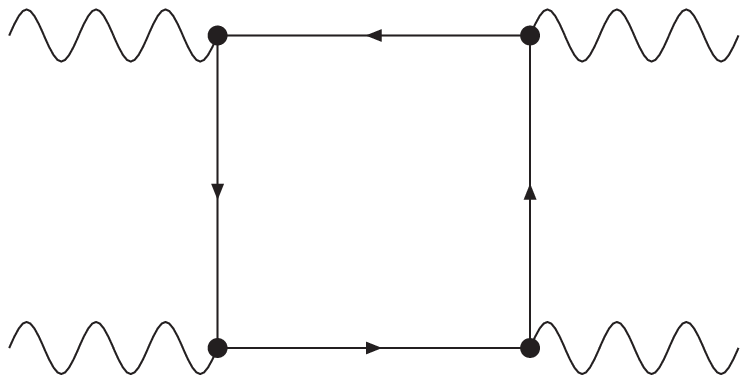}
  \end{minipage} +
  \begin{minipage}{2.2cm}
    \includegraphics[width=2.2cm]{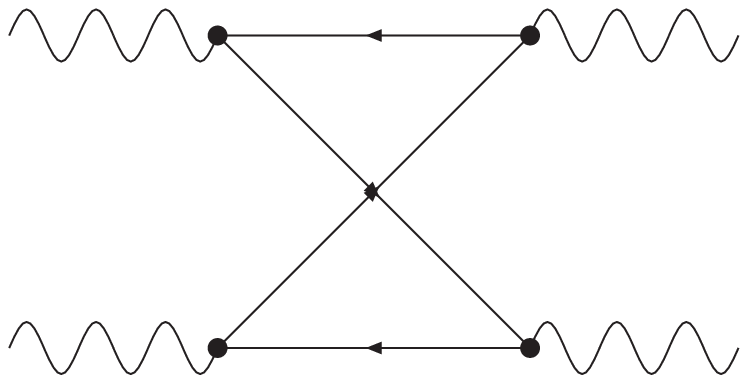}
  \end{minipage} +
  \begin{minipage}{2.2cm}
    \includegraphics[width=2.2cm]{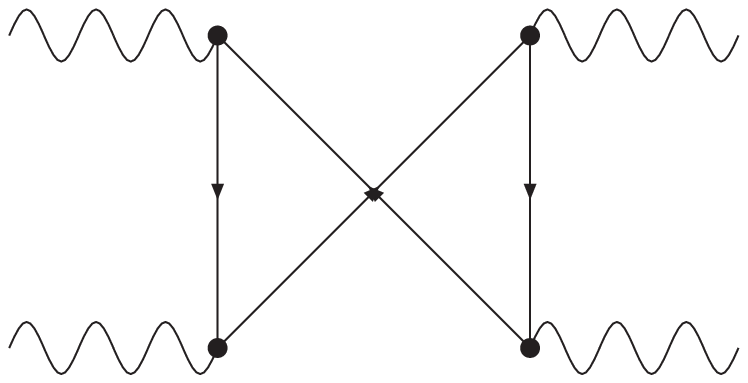}
  \end{minipage} \, .\nonumber
\end{eqnarray}
\caption{Lowest order diagrams}
  \label{fig:boxdiafer}
\end{figure}

The lowest order diagrams\footnote{We perform all computations in
the Feynman gauge.} in Fig.\,\ref{fig:boxdiafer}
contributing to the correlation function $A$
are fermionic boxes (one-loop).
At high energies, they behave as $\log^2 s$ \cite{Bartels:2003dj}, and 
therefore give a contribution
to the total cross section which decreases as $1/s$.
Radiative gluonic corrections to these fermion loop 
graphs will not modify the power of the energy dependence but provide 
double logarithmic corrections.

\subsection{Two gluon exchange}

\begin{figure}[htbp]
  \includegraphics[width=4.2cm]{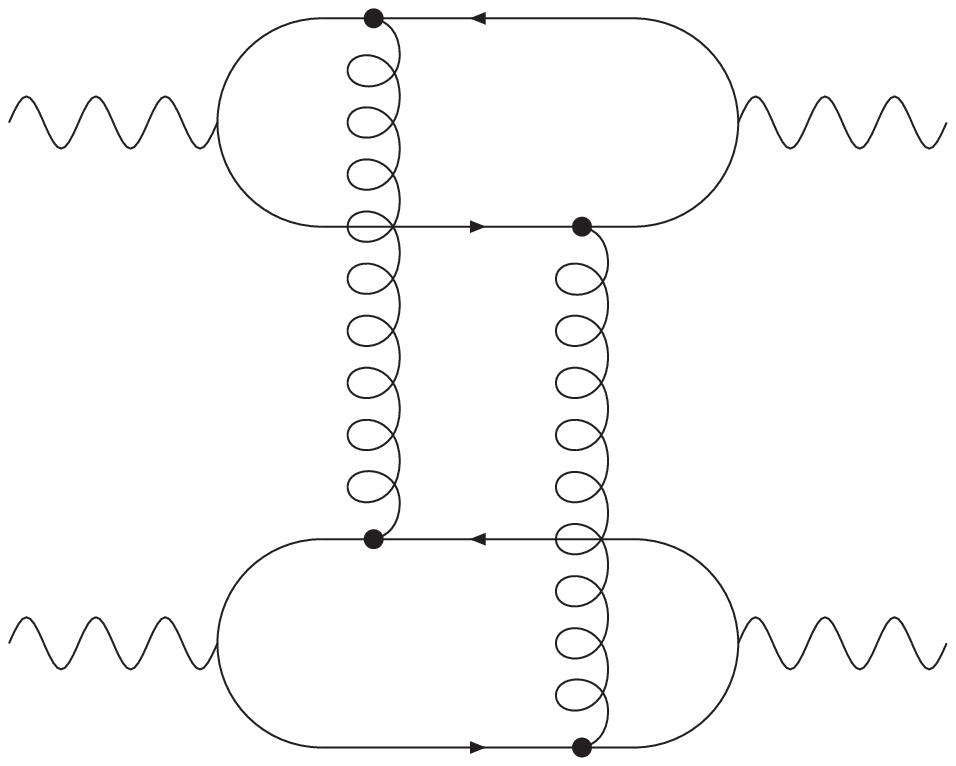}
  \includegraphics[width=4.2cm]{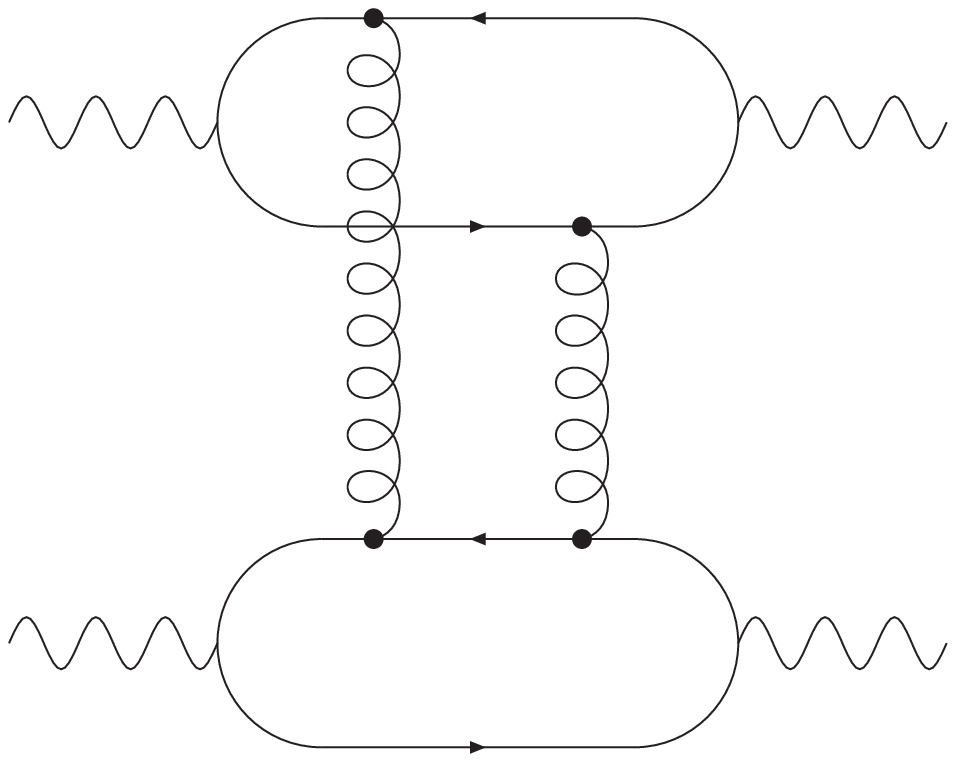}
  \caption{An example of three-loop diagrams corresponding to two gluon exchange.}
  \label{fig:twoGluonsEx}
\end{figure}

\begin{figure}[htbp]
  \includegraphics[width=4.2cm]{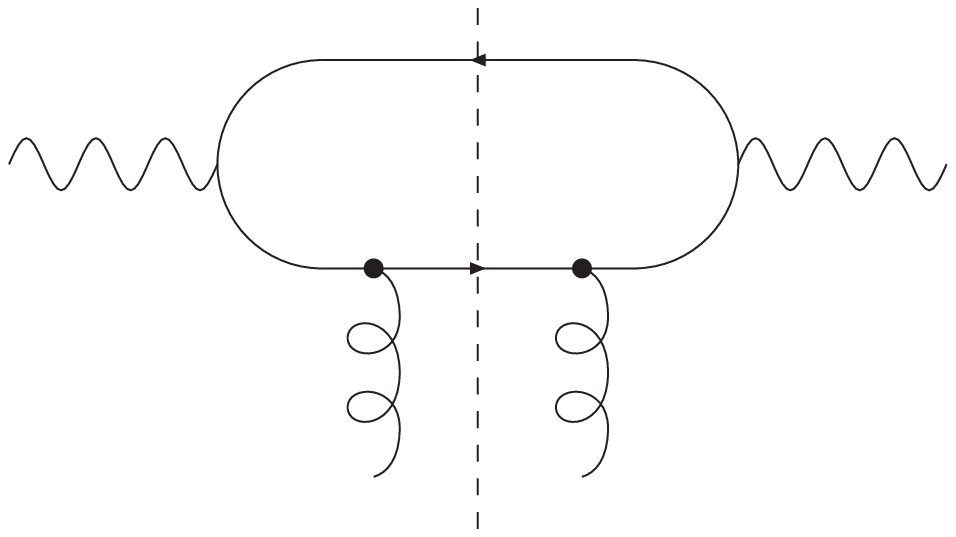}
  \includegraphics[width=4.2cm]{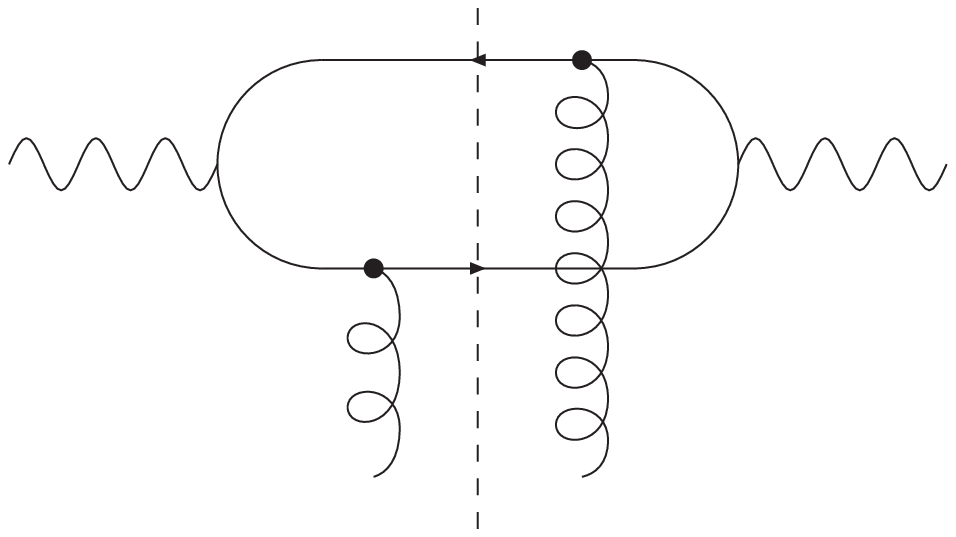}\\[20pt]
  \includegraphics[width=4.2cm]{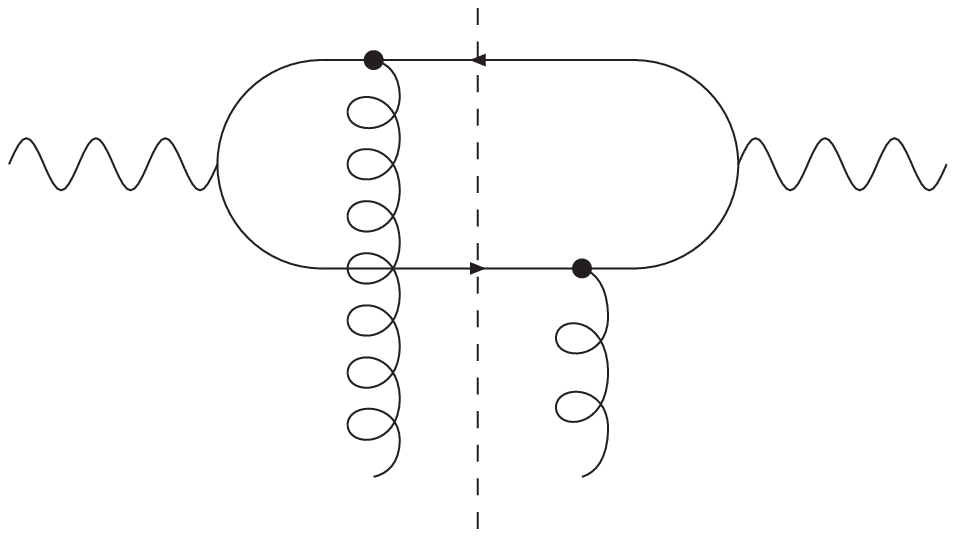}
  \includegraphics[width=4.2cm]{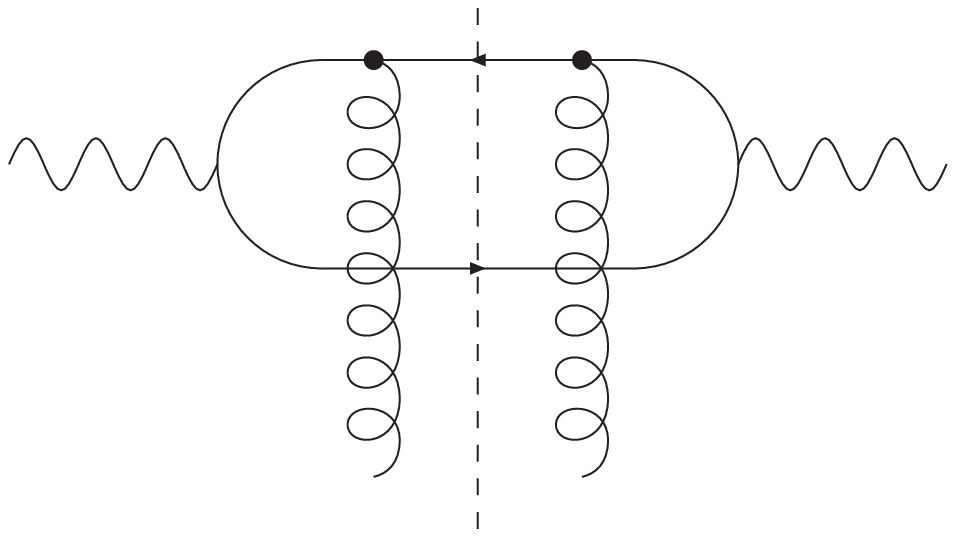}
  \caption{The one-loop diagrams contributing to the impact factor $\Phi$.}
  \label{fig:IFQCD}
\end{figure}

At the three-loop level a new class of diagrams
becomes available, in which purely gluonic $t$-channel states appear. As an example, two lowest order 
diagrams are shown in Fig.\ \ref{fig:twoGluonsEx}.
At high energies the sum of all lowest order diagrams, $A$, behaves as $\alpha_s^2 s$, and
therefore provides a contribution to the total cross section which (up to powers 
of $\ln s$) is constant in $s$.
It is clear that at high energy, independently of how small
$\alpha_s$ is, these diagrams dominate with respect to the boxes
and their radiative corrections. 
In the Regge limit the lowest order 
diagram, $A^{(0)}$, is purely imaginary and takes the form 
\begin{eqnarray}
  \label{eq:AzeroFactorized}
  &&A^{(0)}(s,t) =\\
&& i s \int \frac{d^2 \bk}{(2\pi)^2\bk^2 (\bq-\bk)^2}
  \Phi^{a_1 a_2}_A(\bk, \bq-\bk)\Phi^{a_1 a_2}_B(\bk, \bq-\bk) \, .\nonumber
\end{eqnarray}
Here the so called \emph{impact factors} $\Phi$ (Fig.\ \ref{fig:IFQCD}) represent the coupling of the 
virtual photons to the two $t$-channel gluons. Their precise definition is
\begin{eqnarray}
  \label{eq:IFdefinition}
  &&\Phi^{\lambda_A \lambda_{A'} a a'}_A(\bk, \bk') = \frac{1}{s^2}  \epsilon^{\lambda_A}_{\mu_A}(p_A)^* \epsilon^{\lambda_{A'}}_{\mu_{A'}}(p_{A'})~  p_{2 \rho} p_{2 \rho'}\nonumber\\
  &&\qquad\times\int \frac{ds_1}{2\pi}
  \Im A^{\mu_A \mu_{A'} \rho \rho'}_{\gamma q \to \gamma q} (s_1,t)
\end{eqnarray}
with a similar definition for $\Phi_B$.
Here $\Im A^{\mu_A \mu_{A'} \rho \rho'}_{\gamma q \to \gamma q} (s_1,t)$
is the imaginary part of the amplitude for the scattering of the
virtual photon $A$ with polarization $\lambda_A$ and a gluon
with momentum $-k$, Lorenz index $\rho$,
and color label $a$ into the photon $A'$ with polarization $\lambda_{A'}$
and a gluon with momentum $k$, Lorenz index $\rho'$, and color label $a'$.
$s_1$ is the total energy squared of the photon-gluon system,
and it is related to the Sudakov component $\beta$ of $k$
(which in this regime is the same as the one of $k'$)
along the Sudakov vector $p_2$ by
$s_1=(p_A-k)^2 \simeq -Q_A^2 -\bk^2 - s \beta \approx -s \beta$.
For each $t$-channel gluon, we have a factor $2p_{2 \rho} p_{1 \sigma}/s$,
since, in the Regge limit, only a specific component of the gluon polarization
tensor contributes to the leading power in $s$, namely,
\begin{equation}
  \label{eq:gluonPolLeading}
  g_{\rho \sigma} = \frac{2}{s}\big(p_{2 \rho}  p_{1 \sigma} + 
  p_{1 \rho} p_{2 \sigma} \big) +
  g_{\perp \rho \sigma} \to
  \frac{2}{s} p_{2 \rho} p_{1 \sigma}  \, .
\end{equation}
With these definitions the impact factors
$\Phi_{A,B}$ are independent of $s$. They depend, in the leading approximation
we are interested in, only upon the virtuality and the polarizations of the photons,
the gluon colors, and the transverse momenta.

\subsection{All-order summation in the leading logarithmic approximation}
\label{sec:LL}
Generalizing, in the leading logarithmic approximation, the lowest order diagrams 
to higher orders in $\alpha_s$, the two gluon exchange is replaced by the BFKL 
\cite{Kuraev:1976ge-Kuraev:1977fs-Balitsky:1978ic}
Green's function:  
\begin{equation}
  \label{eq:AResummed}
  A(s,t) = i s~ \Phi_A \otimes G(s) \otimes \Phi_B \, ,  
\end{equation}
where we have introduced the symbol $\otimes$ to denote the
transverse momentum convolution of \eqref{eq:AzeroFactorized}, including
the transverse gluon propagators and the contraction of the color indices.
$G(s)$ is the Green's function of the BFKL equation,
accounting for the resummed LL corrections.
The LLA sums the radiative corrections to $A^{(0)}$ in \eqref{eq:AzeroFactorized}, and it 
is valid in the region where $\alpha_s \ll 1$ and \mbox{$\alpha_s \log s \sim 1$}.
The BFKL Pomeron denotes the bound state formed by two interacting Reggeized
gluons with the quantum numbers of the vacuum (for more details see, for example,
\cite{Lipatov:1989bs,Lipatov:1996ts,Forshaw:1997dc}
and references therein). In LLA, the BFKL Green's function contains only gluonic 
contributions; fermionic corrections appear only in the next-to-leading correction.
As a consequence of this, when turning to the supersymmetric extension of QCD, 
the LLA of the BFKL Pomeron remains the same as in QCD. 
What needs to be studied is 
the role of the scalar degrees of freedom in the box diagrams and in the 
impact factors. This will be done in the following section.

\section{$\cN=4$ SYM and $R$-currents}

The maximally supersymmetric non-Abelian gauge theory in four dimensions
admits $\cN=4$ supersymmetries. It contains a vector multiplet in
the adjoint representation of the gauge group \SUN. The theory
enjoys a \SUR~ global symmetry, called $R$-symmetry, which transforms 
the different supercharges. In terms of component fields the theory has
\begin{itemize}
\item
  1 vector field $A_\mu$, scalar of \SUR;
\item
  4 chiral spinors $\lambda_I$ in the fundamental representation of \SUR;
\item
  6 real scalars $X_M$ in the vector representation of \SUR.
\end{itemize}
Capital indices transform under the $R$-symmetry group. In particular,
$A,B,C,...=1,...,15$ are indices of the adjoint representation,
$I,J,K,...=1,...,4$ transform under the fundamental, and
$M,N,...=1,...,6$ under the vector representations of the $R$-symmetry.
Small indices $a,b,c,...=1,...,N_c^2-1$ are adjoint representation
indices for the gauge group \SUN.
Since all the fields live in the adjoint representation of \SUN,
we can write $\Phi=\Phi_{ab}=\Phi^c (t^c)_{ab}$, with $(t^c)_{ab}=-i f^{acb}$ with
$f^{abc}$ being the \SUN~ structure constants, $[t^a,t^b]=i f^{abc} t^c$. Our convention for the normalization of the generators $t^a$ is such that $\tr(t^at^b)=\delta^{ab}/2$.

The Lagrangian is \cite{Brink:1976bc}
\begin{eqnarray}
  \label{eq:LSYM4}
    \cL &=& \tr\Big(
    - \frac{1}{2} F_{\mu\nu}F^{\mu\nu}
    +  D_\mu X_M D^\mu X_M
    + 2 i \lambda_I \sigma^\mu D_\mu \lambdab^I \nonumber\\
    &&-2 i g \lambda_I [\lambda_J,X^{IJ}]
    - 2 i g \lambdab^I [\lambdab^J,X_{IJ}]\nonumber\\
   &&+ \frac{1}{2} g^2 [X_M,X_N][X_M,X_N]
    \Big),
\end{eqnarray}
where $X_M$ and $X_{IJ}$ are related by the $SU(4)\cong SO(6)$
sigma symbols:
\begin{equation}
  \label{eq:SigmaSymb}
  X_{IJ} = -\frac{1}{2} (\Sigma_M)_{IJ} X_M \, ,\qquad
  X^{IJ} = \frac{1}{2} (\Sigma^{-1}_M)^{IJ} X_M \, ,
\end{equation}
with $\Tr(\Sigma_M\Sigma_N^{-1}) = 4 \delta_{MN}$, which implies that
\,\mbox{$X_M X_M = X_{IJ} X^{IJ}$}.
The covariant derivative $D_\mu$ and the gauge field strength tensor
$F_{\mu\nu}$ are defined as usual by\footnote{With $\Phi$ we denote any field
in the theory, $X$, or $\lambda$.}
\begin{eqnarray}
  \label{eq:DmuFmunu}
  D_\mu \Phi &=& \partial_\mu \Phi - i g [A_\mu, \Phi] \, ,\\
  F_{\mu\nu} &=& \partial_\mu A_\nu - \partial_\nu A_\mu
                 - i g [A_\mu, A_\nu] \, .
\end{eqnarray}

\subsection{$R$-symmetry currents and the four point function}

The Lagrangian \eqref{eq:LSYM4} is invariant under the global
transformation ($R$-symmetry)
\begin{equation}
  \label{eq:Rsymtransf}
  \Bigg\{ \begin{array}{lcl}
    \delta \lambda^{a \alpha I} & = & i \epsilon_A
    \lambda^{a \alpha J} (T^A)_{JI} \, ,\\
    \delta \lambdab^{a \dot{\alpha} I} & = & - i \epsilon_A
    (T^A)_{IJ} \lambdab^{a \dot{\alpha} J} \, ,\\
    \delta X^a_M & = & i \epsilon_A (T^A)_{M N} X^a_N \, ,
  \end{array}
\end{equation}
where $\epsilon_A$ are small parameters, and $T^A$ are the \SUR~
generators in the appropriate representation.

The Noether current of the symmetry is
\begin{equation}
  \label{eq:Rcurrent}
  J_R^{\mu A} =
  i \frac{\partial \cL}{\partial(\partial_\mu \Phi)} \Delta^A \Phi =
  \tr \Big(
  - \lambda \sigma^\mu T^A \lambdab
  - i  X T^A D^\mu X \Big) \, ,
\end{equation}
where $\Delta^A \Phi$ is obtained from \eqref{eq:Rsymtransf} with the definition
$\delta\Phi = i \epsilon_A \Delta^A \Phi$ for an infinitesimal
$R$-transformation.

We are interested in evaluating the momentum space
four point function defined in analogy to \eqref{eq:4pointfuncQCD},
\begin{eqnarray}
  \label{eq:4pointfunc}
    &&i (2\pi)^4 \delta(p_A+p_B-p_{A'}-p_{B'}) A_R(s,t)^{\mu_A \mu_B \mu_{A'} \mu_{B'}} =\nonumber\\
    &&\int \prod_i d^4x_i~
    e^{- i p_A \cdot x_A - i p_B \cdot x_B
      +i p_{A'} \cdot x_{A'} + i p_{B'} \cdot x_{B'} }\nonumber\\
&&\times\langle J_R^{A\mu_A}(x_A) J_R^{B\mu _B}(x_B) J_R^{A'\mu_{A'}}(x_A') J_R^{B'\mu_{B'}}(x_B') \rangle
\end{eqnarray}
at weak coupling in the Regge limit \eqref{eq:ReggeLim}.

\subsection{Ward identities}
From \eqref{eq:WardId1} and \eqref{eq:commJ0O} we can compute explicitly
the Ward identities satisfied by \eqref{eq:4pointfunc}. We only need
to specialize \eqref{eq:commJ0O} to the case of interest:
\begin{equation}
  \label{eq:commJA0JB}
  [J_R^{A0}(\vec{x},t),J_R^{B\mu}(\vec{y},t)] =
  \delta^{(3)}(\vec{x}-\vec{y}) (T^A)^B_C J_R^{C\mu}(\vec{x},t) \, .
\end{equation}
The nonvanishing of the commutators \eqref{eq:commJA0JB}, which is due to the
fact that conserved currents of a non-Abelian symmetry are charged,
implies immediately that also the contact terms in the Ward identities
do not vanish,
\begin{eqnarray}
    \label{eq:WardId5}
    &&\partial_\mu \langle J_R^{A\mu}(x) J_R^{A_1\mu_1}(x_1)...
    J_R^{A_n\mu_n}(x_n) \rangle =\sum_{i=1}^n \delta^{(4)} (x-x_i)\nonumber\\
   &&\times \langle J_R^{A_1\mu_1}(x_1)...(T^A)^{A_i}_C J_R^{C\mu_i}(x)...
    J_R^{A_n\mu_n}(x_n) \rangle \, .
\end{eqnarray}
Compared to the QCD case, this introduces some additional complications. In particular,
the standard computation according to which the four point function is finite,
despite n\"aive power counting which suggests a logarithmic divergence,
does not apply anymore. Explicit computation shows that
the UV poles still cancel, but now as a result of the interplay between
the scalar and fermionic sectors (Sec. \ref{sec:oneLoop}).
It is therefore a consequence of supersymmetry.

The change of the Ward identities, at first sight, also affects 
our use of the polarization vectors. 
The simplifications
which lead from (\ref{eq:polVecExplicitLA}-\ref{eq:polVecExplicithApBp})
to (\ref{eq:polVecReducedLAAp}-\ref{eq:polVecReducedhBp})
were only possible  because of the simple Ward identities \eqref{eq:WardId4}, and 
the more complicated identities (\ref{eq:WardId5}) spoil this argument.   
If, however, instead of the full \SUR\, group we restrict
ourselves to a $U(1)$ subgroup of \SUR, we reach a situation similar to the QCD case.
Restriction to the $U(1)$\footnote{Following \cite{CaronHuot:2006te} we choose a 
particular linear combination of the three diagonal \SUR~ generators. In the following we 
will drop the \SUR~ label $A$ in $J_R$ for the $U(1)$ current.} 
means that, on the RHS of (\ref{eq:WardId5}), all $(T^A)^{A_i}_C=-if^{AA_iC}$ 
vanish, and one recovers the same Ward identities
without contact terms as in QCD:
\begin{equation}
  \label{eq:WardId6}
  \partial_\mu \langle J_R^{\mu}(x) J_R^{\mu_1}(x_1)...J_R^{\mu_n}(x_n)
  \rangle = 0 \, .
\end{equation}
We therefore can proceed as before and, via Eq. \eqref{eq:A4decompQCD}, 
conveniently compute projections of $A_R$ onto specific polarization vectors. 

\subsection{An excursion into scalar QED}
\label{sec:ScalarQED}
\begin{figure} 
\begin{equation*}
  A_{2\mu \nu}^{(0)}(x,y) =
  \begin{minipage}{5cm}
    \includegraphics[width=5cm]{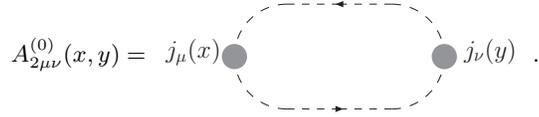}
  \end{minipage} \, .
\end{equation*}
\caption{Lowest order diagram contributing to the two point function of the electromagnetic current.}
\label{eq:A20diagram1Pic}
\end{figure}

As we mentioned already at the end of Sec. \ref{sec:WI},
working in Fourier space requires some care with renormalization. The problem can be
easily illustrated in the simple framework of scalar QED.
Let us consider, as an example, the two point function
$A_2 = \langle j_\mu(x) j_\nu(y) \rangle$
of the electromagnetic current
$j_\mu = i(\varphi \partial_\mu \varphi^* - \varphi^* \partial_\mu \varphi)
-2 e A_\mu \varphi^* \varphi$.
Again, $U(1)$ gauge symmetry implies the Ward identity
$\partial_\mu \langle j_\mu(x) j_\nu(y) \rangle = 0$.
The computation of the lowest order in perturbation theory
performed in the coordinate space with $x \ne y$,
would involve just one  diagram, Fig.\,\ref{eq:A20diagram1Pic}.
Going to momentum space, this diagram gives
\begin{equation}
  \label{eq:A20diagram1}
  A_{2\mu \nu}^{(0)}(p) =
  \int \frac{d^dk}{(2\pi)^d} \frac{(2k-p)_\mu(2k-p)_\nu}{k^2(k-p)^2} \, .
\end{equation}
One immediately sees that \eqref{eq:A20diagram1} does not satisfy
the Ward identity,
\begin{equation}
  \label{eq:WISQED}
  p^\mu A_{2\mu \nu}^{(0)}(p) =
  2 p_\nu \int \frac{d^dk}{(2\pi)^d} \frac{1}{k^2} \, .
\end{equation}
The problem arises because the product of two currents in $A_2$ is not
regular when $x \to y$. As it is well known, such a product is defined
by an operator product expansion (OPE),
\begin{equation}
  \label{eq:jjOPE}
  j_\mu(x) j_\nu(y) \mathop {\longrightarrow }\limits_{x \to y}
  {\textstyle\sum_i} C_i \frac{\cO_i(y)}{(x-y)^{\alpha_i}} \, ,
\end{equation}
where $\cO_i$ are operators with the same quantum numbers as $j_\mu j_\nu$.
Equation \eqref{eq:jjOPE} means that the product
of two currents at the same point mixes with the operators $\cO_i$.
By dimensional analysis it is easy to spot the operator which, in \eqref{eq:jjOPE}, 
gives the leading singularity:
\begin{equation}
  \label{eq:jjOPEleading}
  j_\mu(x) j_\nu(y) =
  C~ \frac{(g_{\mu\nu}\varphi^*\varphi)(y)}{(x-y)^d} + \cO\big( (x-y)^{-d+2} \big) \, .
\end{equation}
Note that such singular behavior is precisely on the boundary of convergence
of the Fourier integrals, and all the other terms in the OPE contain 
integrable singularities. It is therefore enough to regularize the divergence 
by removing this leading term:
\begin{equation}
  \label{eq:newJJ}
  j_\mu(x) j_\nu(y) \to
  j_\mu(x) j_\nu(y) - C~ \frac{(g_{\mu\nu}\varphi^*\varphi)(y)}{(x-y)^d} \, .
\end{equation}
In momentum space the operator $g_{\mu\nu}\varphi^*\varphi$ leads, at the one-loop
level, to an additional diagram..
\begin{eqnarray}\label{eq:A20diagram2Pic}
  &&A_{2\mu \nu}^{(0)}[g_{\mu\nu}\varphi^*\varphi](x,y)=
  \begin{minipage}{2.5cm}
    \includegraphics[width=2.5cm]{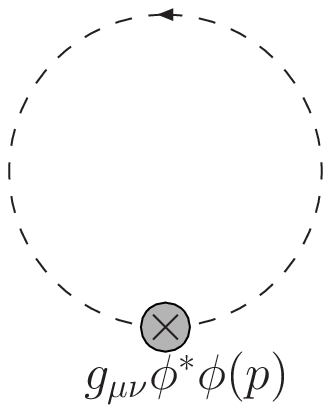}
  \end{minipage}\nonumber\\
  &&\qquad= - C g_{\mu\nu} \int \frac{d^dk}{(2\pi)^d} \frac{1}{k^2} \, .
\label{fig:A20diagram2Pic}
\end{eqnarray}

Comparing \eqref{eq:A20diagram2Pic} with \eqref{eq:WISQED} we can fix
$C=2$. With this we obtain, for $A_{2\mu \nu}^{(0)}$, the same result
that one would get from computing the one-loop correction of the photon
self energy $\langle A_\mu A_\nu \rangle$ in scalar QED, which indeed
satisfies the Ward identity $p^\mu A_{2\mu \nu}^{(0)}(p) = 0$.

The same argument applies to the scalar sector of the $\cN=4$ theory.
One has to add the appropriate regularizing diagrams, which ensure that the correlation
functions are well defined and fulfill the Ward identities in
momentum space.

Before returning to the $\cN=4$ theory, we observe that the QED computation
we just sketched would simplify considerably if we were interested only in
the imaginary part of $A_2$. The term in \eqref{eq:A20diagram2Pic}
is real and does not contribute, while the imaginary part of
\eqref{eq:A20diagram1}, which is easily computed by means of the cutting rules,
now fulfills the Ward identity, thanks to the delta-functions of the 
two on-shell scalar propagators.

\subsection{One-loop diagrams}
\label{sec:oneLoop}
An important step in checking the Regge factorization of the $R$-current
scattering amplitude is to verify that the fermionic and scalar one-loop diagrams
are subleading at high energies. This task includes the correct 
regularization of ultraviolet divergencies. For correlators of $R$-currents which 
belong to a $U(1)$ subgroup, we will show that this task can be solved by applying the 
previous arguments. 
When considering a correlation function with arbitrary \SUR \, labels
the situation is not as simple as in QCD or scalar QED.
The usual argument for the absence of UV divergencies is based on the Ward 
identities \eqref{eq:WardId4} and does not work in the present case.
Nevertheless, by performing the explicit computation, we can prove that 
the one-loop diagrams are UV finite. It will be shown that, 
in this situation, it is supersymmetry that constrains the UV divergence 
to be absent. More precisely, it is the interplay between the fermionic and scalar
sectors which leads to cancellations.

\subsubsection*{UV poles}
\begin{figure}[htbp]
  \begin{minipage}{3.2cm}
    \centering
    \includegraphics[width=3.2cm]{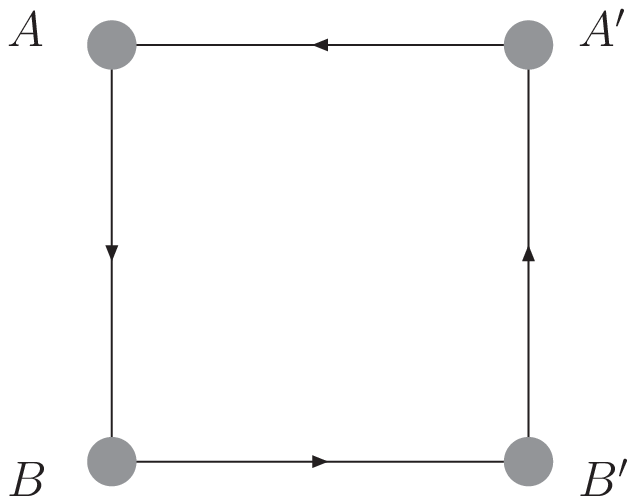}\\
    BF1
  \end{minipage}
  \begin{minipage}{3.2cm}
    \centering
    \includegraphics[width=3.2cm]{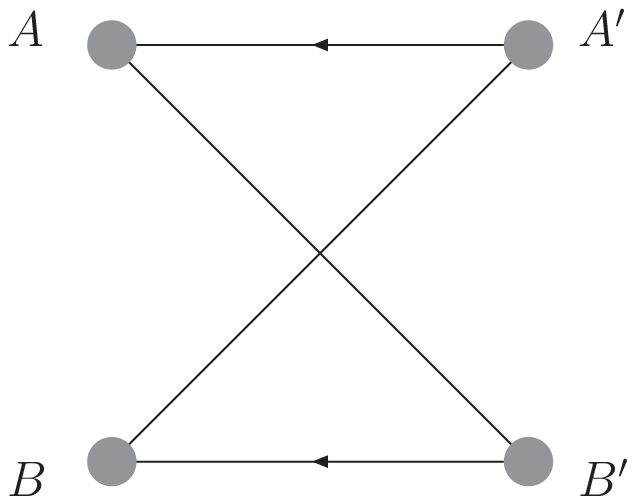}\\
    BF2
  \end{minipage}
  \begin{minipage}{3.2cm}
    \centering
    \includegraphics[width=3.2cm]{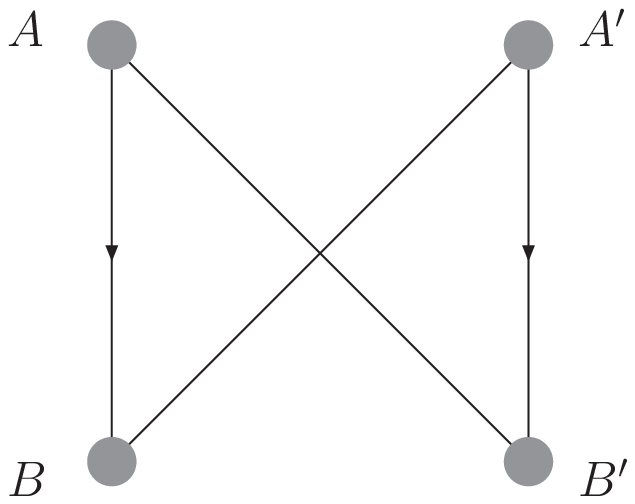}\\
    BF3
  \end{minipage}
  \caption{One-loop diagrams with fermions}
  \label{fig:FBoxes}
\end{figure}

The one-loop fermionic diagrams are the same boxes as in QCD, depicted in Fig.\
\ref{fig:FBoxes}.
In order to discuss their UV behavior, we regularize the IR
region by giving the fermions a small mass $m$.
The UV singularities of the diagrams $BF1-3$ can be easily
computed:
\begin{eqnarray}
  \label{eq:BF1UV}
   && BF1_{UV}=
    \frac{2}{3} \frac{i \pi^{2-\epsilon}m^{-2\epsilon}}{(2\pi)^4} \Gamma(\epsilon)
    \Tr \Big( T^A T^{A'} T^{B'} T^B \Big) \\
    &&\times\Big( g_{\mu_A \mu_{A'}} g_{\mu_B \mu_{B'}} + g_{\mu_A \mu_B} g_{\mu_{A'} \mu_{B'}} -
    2 g_{\mu_A \mu_{B'}} g_{\mu_{A'} \mu_B} \Big)\,.\nonumber
\end{eqnarray}
The contributions $BF2_{UV}$ and $BF3_{UV}$ can be obtained from \eqref{eq:BF1UV}
by permuting indices. It is immediately clear that their sum does not vanish
unless we restrict ourselves to the $U(1)$ subgroup, and all the \SUR \ traces  
are the same. In this case the cancellation
works precisely as in QCD.

\begin{figure*}[htbp]
  \centering
  \begin{minipage}{3cm}
    \centering
    \includegraphics[width=3cm]{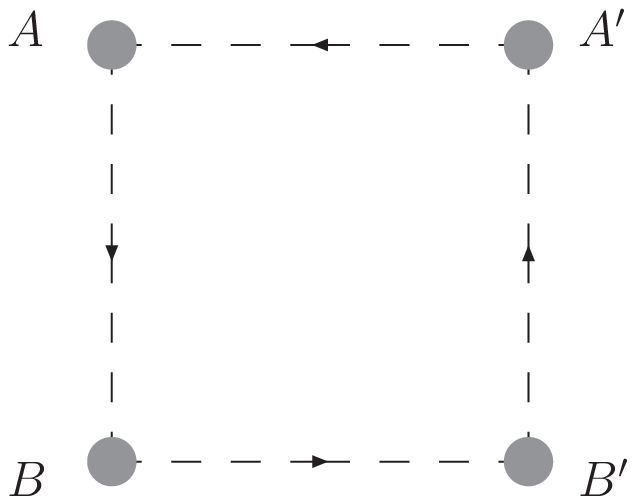}\\
    BS1
  \end{minipage}
  \begin{minipage}{3cm}
    \centering
    \includegraphics[width=3cm]{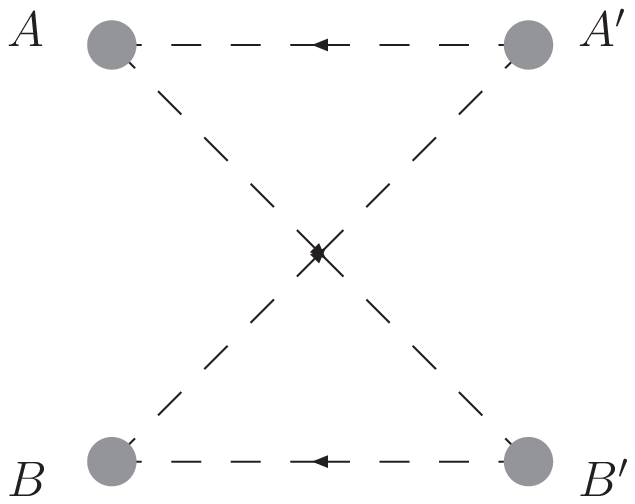}\\
    BS2
  \end{minipage}
  \begin{minipage}{3cm}
    \centering
    \includegraphics[width=3cm]{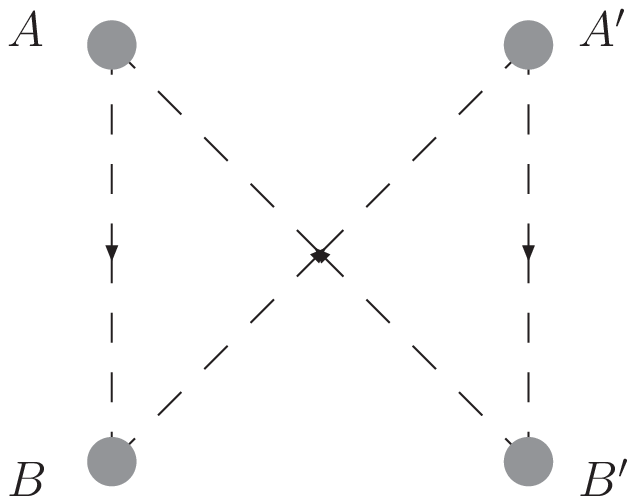}\\
    BS3
  \end{minipage}
  \begin{minipage}{3cm}
    \centering
    \includegraphics[width=3cm]{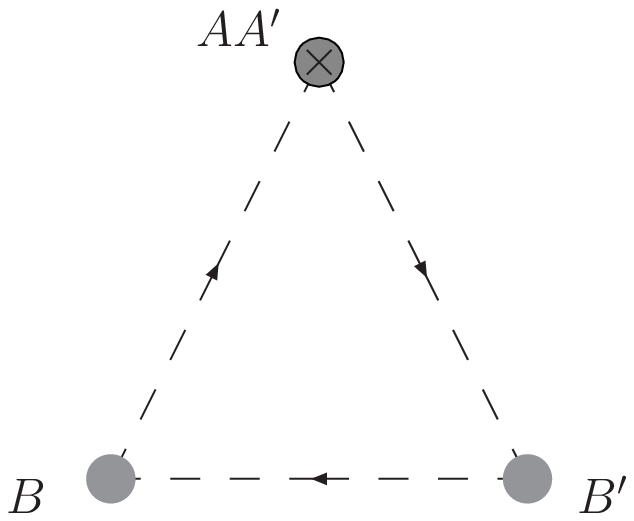}\\
    BS4
  \end{minipage}
  \begin{minipage}{3cm}
    \centering
    \includegraphics[width=3cm]{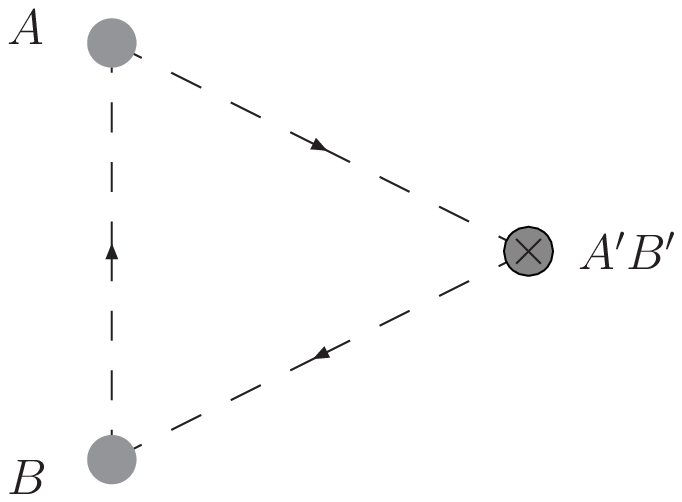}\\
    BS5
  \end{minipage}
  \begin{minipage}{3cm}
    \centering
    \includegraphics[width=3cm]{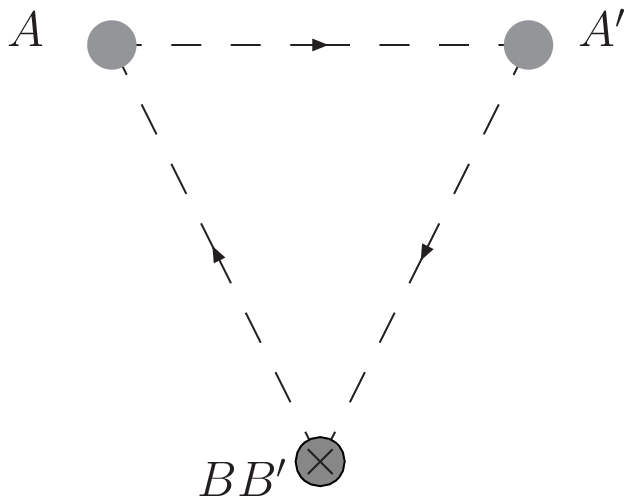}\\
    BS6
  \end{minipage}
  \begin{minipage}{3cm}
    \centering
    \includegraphics[width=3cm]{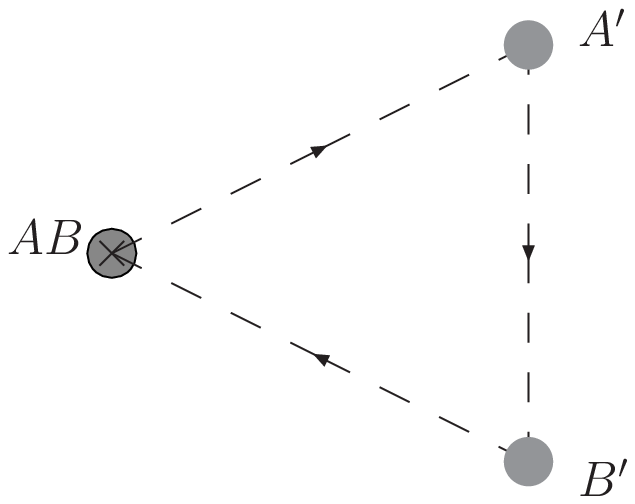}\\
    BS7
  \end{minipage}
  \begin{minipage}{3cm}
    \centering
    \includegraphics[width=3cm]{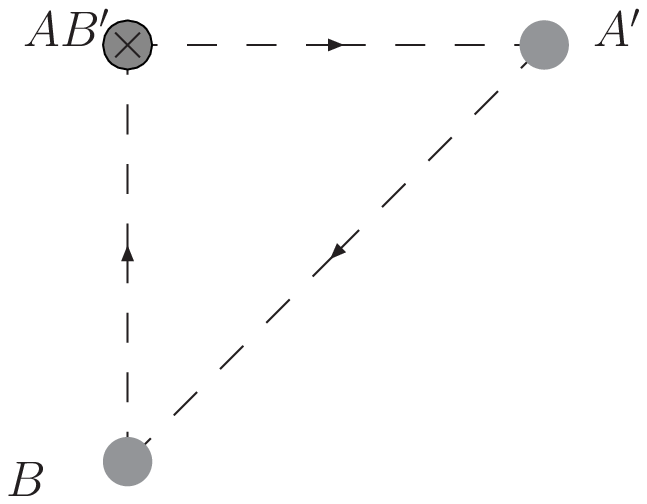}\\
    BS8
  \end{minipage}
  \begin{minipage}{3cm}
    \centering
    \includegraphics[width=3cm]{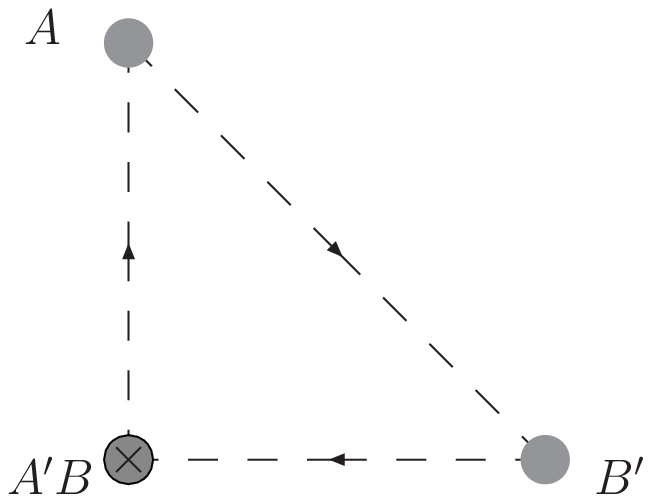}\\
    BS9
  \end{minipage}
  \begin{minipage}{1.8cm}
    \centering
    \includegraphics[width=1.8cm]{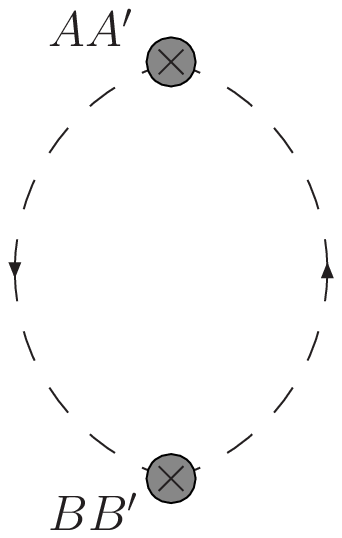}\\
    BS10
  \end{minipage}
  \begin{minipage}{3.8cm}
    \centering
    \includegraphics[width=3.8cm]{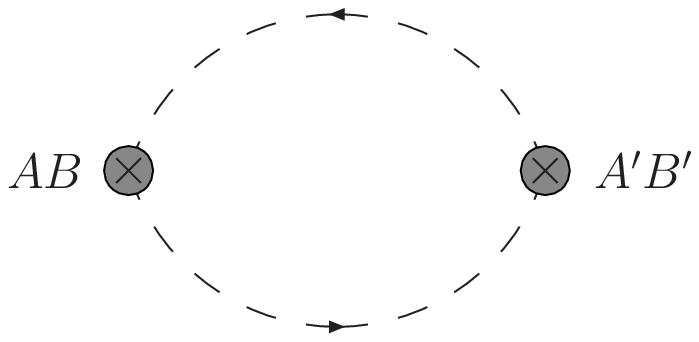}\\
    BS11
  \end{minipage}
  \begin{minipage}{3.6cm}
    \centering
    \includegraphics[width=3.6cm]{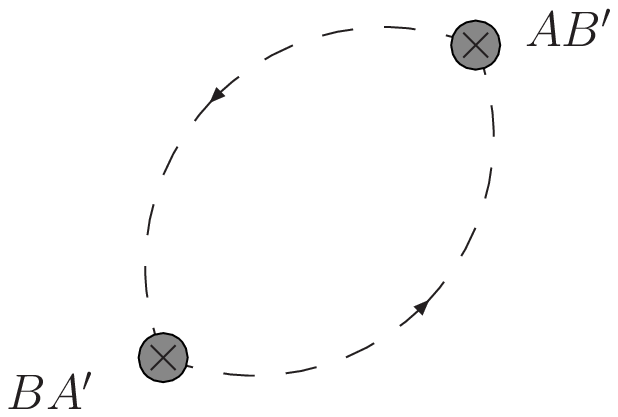}\\
    BS12
  \end{minipage}
  \caption{One-loop diagrams with scalars}
  \label{fig:1loopscalars}
\end{figure*}

There are 12 one-loop scalar diagrams (including those which are required for 
regularization), and they are all depicted in
Fig.\ \ref{fig:1loopscalars}.
From the UV region of these diagrams one obtains
\begin{eqnarray}
  \label{eq:BS1UV}
  &&BS1_{UV}= \frac{2}{3} \frac{i \pi^{2-\epsilon}m^{-2\epsilon}}{(2\pi)^4} \Gamma(\epsilon)\Tr \Big( T^A T^{A'} T^{B'} T^B \Big)\\
  &&\times\Big( g_{\mu_A \mu_{A'}} g_{\mu_B \mu_{B'}} + g_{\mu_A \mu_B} g_{\mu_{A'} \mu_{B'}} +
  g_{\mu_A \mu_{B'}} g_{\mu_{A'} \mu_B} \Big) \, , \nonumber\\
  \label{eq:BS4UV}
  &&BS4_{UV}=
  -2 \frac{i \pi^{2-\epsilon}m^{-2\epsilon}}{(2\pi)^4} \Gamma(\epsilon)\\
 &&\qquad\times \Tr \Big( \frac{T^A T^{A'} + T^{A'} T^A}{2} T^{B'} T^B \Big) g_{\mu_A \mu_{A'}} g_{\mu_B \mu_{B'}} \, ,\nonumber \\
  \label{eq:BS10UV}
  &&BS10_{UV}=
  2 \frac{i \pi^{2-\epsilon}m^{-2\epsilon}}{(2\pi)^4} \Gamma(\epsilon)\\
&&\qquad\times  \Tr \Big( \frac{T^A T^{A'} + T^{A'} T^A}{2} \frac{T^B T^{B'} + T^{B'} T^B}{2} \Big)\nonumber\\
&&\qquad\qquad\qquad\qquad\qquad\qquad\qquad\qquad\times g_{\mu_A \mu_{A'}} g_{\mu_B \mu_{B'}} \, .\nonumber
\end{eqnarray}
All the other diagrams in Fig.\ \ref{fig:1loopscalars} can be obtained by
permutations of the indices.

When we restrict ourselves to a $U(1)$ subgroup of 
\SUR~ (which also ensures \eqref{eq:WardId6} to hold) all the traces in
(\ref{eq:BF1UV}-\ref{eq:BS10UV}) coincide, and the UV poles in the scalar sector 
cancel among themselves.

In the case of different \SUR \, generators in \eqref{eq:BF1UV}-\eqref{eq:BS10UV} 
the cancellation of the UV poles does not work for the scalar and fermionic sectors 
separately. But a straightforward computation shows that the sum of the divergent pieces 
of all scalar diagrams, $BS1-12_{UV}$, is just the opposite of the sum of the 
fermionic divergencies, $BF1-3_{UV}$, and therefore
the full one-loop function is UV finite. This cancellation is a result of the supersymmetry.

\subsubsection*{High energy behavior}
Let us now turn to the calculation of the high energy behavior. 
From now on, we restrict the $R$-currents to a $U(1)$ subgroup. The computation of the
fermion boxes at high energy then is just the same as in QCD. We briefly
recall the argument of \cite{Gorshkov:1966ht} (see also
\cite{Bartels:2003dj}), which shows how
a double log emerges. This will also help to prepare the subsequent
computation of the scalar diagrams.

The double log arises because, in the high energy limit, the fermion numerator produces
a term proportional to $s k^2$.
More precisely, the region of integration where the double
log arises is $Q_i^2,\bq^2 \ll \bk^2 \ll \alpha s, \beta s$ and 
\mbox{$x_i \ll \alpha,\beta \ll 1$}, with $x_i=Q_i^2/s$.
In this region the integral is\footnote{The subscript L means that we are
keeping only the leading term in energy, and we drop the trace over
the \SUR~ structure constant, e.g. $\Tr( T^A T^A T^A T^A)$.}
\begin{eqnarray}
  \label{eq:doubleLog1}
  &&BF1_L=-\frac{s}{2(2\pi)^4} \int d\alpha \int d\beta \int d^2\bk\nonumber\\
  &&\times\frac{s}{(s \alpha \beta -(\bk-\bq)^2 + i\epsilon)(-s\beta + i\epsilon)(s\alpha + i\epsilon)} \, ,
\end{eqnarray}
where one of the propagators ($k^2$) has been canceled by the $k^2$ in the numerator.
Closing the $\beta-$contour below we pick up the pole in the first propagator,
and after a shift in $\bk$ we obtain
\begin{equation}
  \label{eq:doubleLog3}
  BF1_L=-\frac{i}{2(2\pi)^3} \int_x^1 \frac{d\alpha}{\alpha} \int_x^1 \frac{d\beta}{\beta} \int d^2\bk~
  \delta (s \alpha \beta -\bk^2) \, ,
\end{equation}
where $x = Q^2/s \simeq x_i$.
Performing the angular integration and then the $\bk^2$ integral via the
delta-function we arrive at
\begin{equation}
  \label{eq:doubleLog4}
  BF1_L=-\frac{i}{4(2\pi)^2} \int_x^1 \frac{d\alpha}{\alpha} \int_{x/\alpha}^1 \frac{d\beta}{\beta} =
  -\frac{i}{8(2\pi)^2} \log^2 \frac{s}{Q^2}\, ,
\end{equation}
which confirms our previous claim about the double log behavior of the
fermion box.

Now we focus on the scalar diagram $BS1$,
\begin{eqnarray}
  \label{eq:BS1}
  BS1&=&  \int \frac{d^4k} {(2\pi)^4}
  \frac{(2k-p_A)_{\mu_A} (2k+p_B)_{\mu_B} } {k^2(k-q)^2}\nonumber\\
&&\times\frac{(2k-q-p_A)_{\mu_{A'}} (2k-q+p_B)_{\mu_{B'} }} {(k-p_A)^2(k+p_B)^2} \, .
\end{eqnarray}
Projecting first onto longitudinal polarizations and keeping only
the leading contribution, we obtain
\begin{eqnarray}
  \label{eq:BS1LL1}
  BS1_L^{LLLL}&=&Q_AQ_{A'}Q_BQ_{B'} \int \frac{d^4k}{(2\pi)^4}\nonumber\\
  &&\times\frac{1}{k^2(k-q)^2(k-p_A)^2(k+p_B)^2} \, ,
\end{eqnarray}
which means that the longitudinal projection reduces simply to the standard
scalar integral we would encounter in a massless $\phi^3$ theory.
It behaves again as a double log, but now the additional logarithm
arises in the infrared region $\bk \simeq 0$ due to the vanishing mass
of the fields. Let us consider indeed the region of integration
$x_i \ll \alpha,\beta \ll 1$ (which we have already used in order to
get to \eqref{eq:BS1LL1}) but $\bk^2 \ll Q_i^2,\bq^2$. There \eqref{eq:BS1LL1}
becomes
\begin{eqnarray}
  \label{eq:BS1LL2}
  &&BS1_L^{LLLL}=\frac{s Q_AQ_{A'}Q_BQ_{B'}}{2(2\pi)^4} \int d\alpha \int d\beta \int d^2\bk\nonumber\\
  &&\quad\quad\times\frac{1}{(s \alpha \beta -\bk^2 + i\epsilon)
    (-\bq^2)
    (-s\beta + i\epsilon)(s\alpha)} \, .
\end{eqnarray}
Again we close the $\beta-$contour below and pick up the pole
from the first propagator, which is now $k^2$
\begin{eqnarray}
  \label{eq:BS1LL3}
  BS1_L^{LLLL} &=& -\frac{i Q_AQ_{A'}Q_BQ_{B'}}{4(2\pi)^2 s \bq^2}\nonumber\\
  &&\times\int_x^1 \frac{d\alpha}{\alpha} \int_x^1 \frac{d\beta}{\beta}
  \int_{\bk^2 \ll Q^2} d\bk^2 \delta( s\alpha\beta - \bk^2 ) \nonumber\\
  &=&  -\frac{i Q_AQ_{A'}Q_BQ_{B'}}{4(2\pi)^2 s \bq^2}
  \int_x^1 \frac{d\alpha}{\alpha} \int_x^{x/\alpha} \frac{d\beta}{\beta}\nonumber \\
  &\simeq&  \frac{i}{8(2\pi)^2} \frac{Q^2}{s} \log^2 \frac{s}{Q^2}\, . 
\end{eqnarray}
Let us consider now transverse polarization. Projecting \eqref{eq:BS1} onto
the transverse polarization (\ref{eq:polVecReducedhAB}-\ref{eq:polVecReducedhBp})
and keeping only leading terms in the numerator, we obtain
\begin{eqnarray}
  \label{eq:BS1TTTT}
  &&BS1_L^{TTTT} =  16 \int d^4k
  \frac{k \cdot \epsilon^{h_A} k \cdot \epsilon^{h_{A'}}
    k \cdot \epsilon^{h_B} k \cdot \epsilon^{h_{B'}}
  }{k^2(k-q)^2(k-p_A)^2(k+p_B)^2} \nonumber\\[10pt]
  &&=
  \big(
  \delta^{h_A h_{A'}} \delta^{h_B h_{B'}} + \delta^{h_A h_B} \delta^{h_{A'} h_{B'}} +
  \delta^{h_A h_{B'}} \delta^{h_{A'} h_B}
  \big)\nonumber\\
  &&\qquad\qquad \times\frac{s}{3(2\pi)^4} \int d\alpha \int d\beta \int d^2\bk\nonumber\\
  &&\qquad\qquad\times\frac{s \alpha \beta - \bk^2}{
    (s \alpha \beta -(\bk-\bq)^2 + i\epsilon)
    (-s\beta + i\epsilon)(s\alpha)} \, .
\end{eqnarray}
As we did in \eqref{eq:doubleLog1} we close the $\beta-$contour below
and get the residue from the pole in the first propagator,
which, after a shift in $\bk$, gives
\begin{eqnarray}
  \label{eq:BS1TTTT2}
    &&BS1_L^{TTTT} =\nonumber\\
    &&\big(
    \delta^{h_A h_{A'}} \delta^{h_B h_{B'}} + \delta^{h_A h_B} \delta^{h_{A'} h_{B'}} +
    \delta^{h_A h_{B'}} \delta^{h_{A'} h_B}
    \big)\nonumber \\
    &&\qquad\times \frac{i}{3(2\pi)^3s} \int \frac{d\alpha}{\alpha} \int \frac{d\beta}{\beta}
    \int d^2\bk \nonumber\\
&&\qquad\times( s \alpha \beta - \bk^2 - \bq^2 -2 \bk \cdot \bq )
    \delta( s \alpha \beta - \bk^2 ) \, . 
\end{eqnarray}
The scalar product vanishes after angular integration, the combination 
$s\alpha\beta - \bk^2$ is set to $0$ through the delta-function,
and the only term left gives
\begin{eqnarray}
  \label{eq:BS1TTTT3}
  &&  BS1_L^{TTTT} \simeq\nonumber\\
&&   - \big(
  \delta^{h_A h_{A'}} \delta^{h_B h_{B'}} + \delta^{h_A h_B} \delta^{h_{A'} h_{B'}} +
  \delta^{h_A h_{B'}} \delta^{h_{A'} h_B}
  \big)\nonumber\\
&&\qquad\times \frac{i}{6(2\pi)^2} \frac{Q^2}{s} \log^2 \frac{s}{Q^2} \, .
\end{eqnarray}
Similar computations can be performed for all the other diagrams in Fig.\
\ref{fig:1loopscalars}, and the results are similar to the one just outlined.
This completes our derivation of the leading high energy behavior of all the one-loop diagrams of
Fig. \ref{fig:FBoxes} and \ref{fig:1loopscalars}. 

We would like to stress the importance of the region $\bk^2 \sim s$:
at first sight, the numerators in \eqref{eq:doubleLog1} and \eqref{eq:BS1LL3}
seem to lead to an even stronger behavior than the one we have computed.
However, in the limit of large $s$, this region coincides with the UV region which has
been discussed at the beginning of this section.
These leading terms cancel when all the diagrams are summed over, in the same fashion as the
cancellation of the UV poles discussed earlier.

As we will see in the next section, the high energy behavior is dominated by gluon exchange,
and the fermion and the scalar box diagrams provide subleading corrections.
This is to be expected since, once the UV finiteness of the one-loop diagrams has been verified, 
we can apply the spin argument, according to which the exchange of two field quanta of spin $s$ 
leads, in the scattering amplitude, to the high energy behavior $\sim s^{2s-1}$. This implies that 
also higher order diagrams in which the box diagrams in Fig.\ \ref{fig:FBoxes} are ``dressed'', 
for examples, by gluon rungs, will have the same power behavior in $s$, modified by powers of 
$\ln^2 s$ (details can be found in ~\cite{Bartels:2003dj}). 
A similar consideration applies to diagrams obtained by ``dressing'' the scalar loops.     
For the leading high energy behavior we are thus left with gluon exchanges: 
using the spin argument one expects, for the scattering amplitude, the high energy  behavior$\sim s$.

\subsection{Two gluon exchange}

\begin{figure}[htbp]
  \includegraphics[width=7cm]{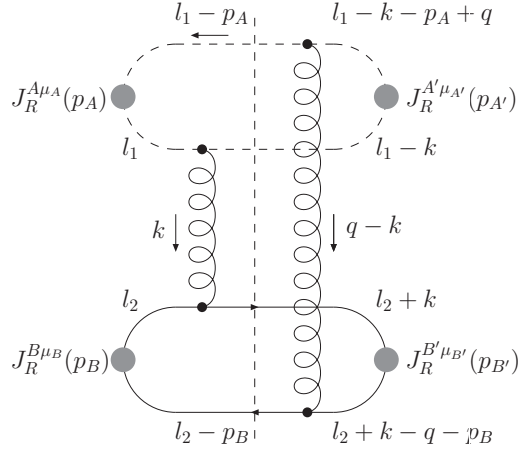}
  \caption{One of the diagrams contributing to the two gluon exchange in $\cN=4$.}
  \label{fig:2gluonsN4}
\end{figure}

As it was the case in QCD, gluon exchange starts at three loops. In Fig.\ \ref{fig:2gluonsN4} we depict one of the 
lowest order diagrams contributing to the two gluon exchange, in order to set the notation for the momenta.
Again, we consider the imaginary part (or, equivalently, the discontinuity in $s$). 
Then we have, in all diagrams, four delta-functions imposing the mass-shell
condition for the intermediate particles (either scalars or fermions).
Two delta-functions are used to fix the integrations over the
longitudinal components of $k$ (the integral in the subenergy $s_1$ in
\eqref{eq:IFdefinition}), and the other two fix one of the two longitudinal integrations
inside each of the impact factors.
The LL contribution arises from the Regge kinematics, in which $\alpha$
is negligible compared to $\alpha_1$, and $\beta$ is negligible compared
to $\beta_2$. Therefore the subdiagrams belonging to the upper impact factor
(scalar loop in Fig.\ \ref{fig:2gluonsN4}) are independent of $\alpha$,
and those of the lower impact factor (the fermion loop) are independent
of $\beta$. This is the mechanism behind the factorization of
\eqref{eq:AzeroFactorized}. In the Regge kinematics also the longitudinal
components of the transverse momentum $q$ are small, $\alpha_q,\beta_q \sim 1/s$,
and dropping them influences only terms suppressed by powers of $s$.

\begin{figure}[htbp]
  \begin{minipage}{2.5cm}
  \centering
    \includegraphics[width=2.5cm]{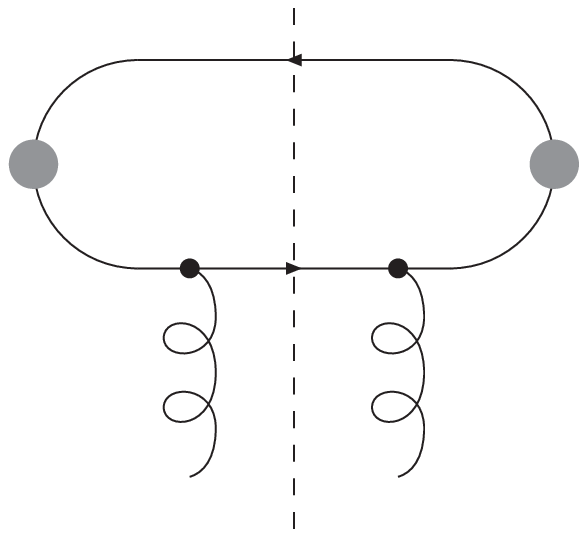}\\
    F1
  \end{minipage}\qquad
  \begin{minipage}{2.5cm}
  \centering
    \includegraphics[width=2.5cm]{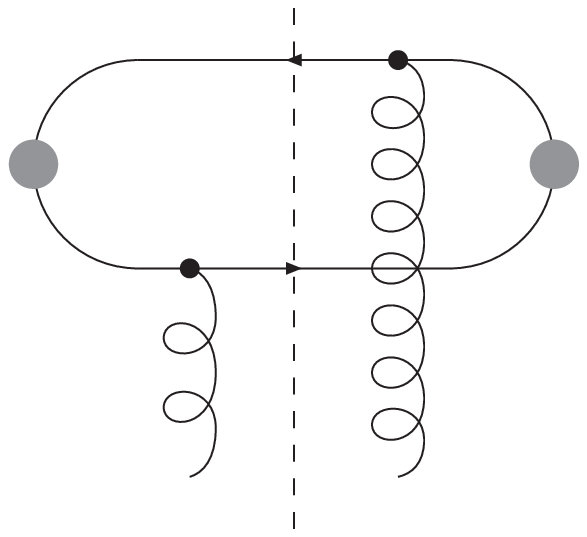}\\
    F2
  \end{minipage}\qquad
  \begin{minipage}{2.5cm}
  \centering
    \includegraphics[width=2.5cm]{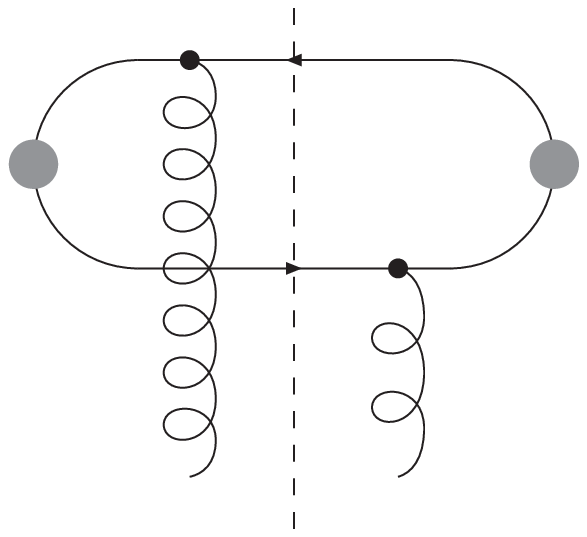}\\
    F3
  \end{minipage}\qquad
  \begin{minipage}{2.5cm}
  \centering
    \includegraphics[width=2.5cm]{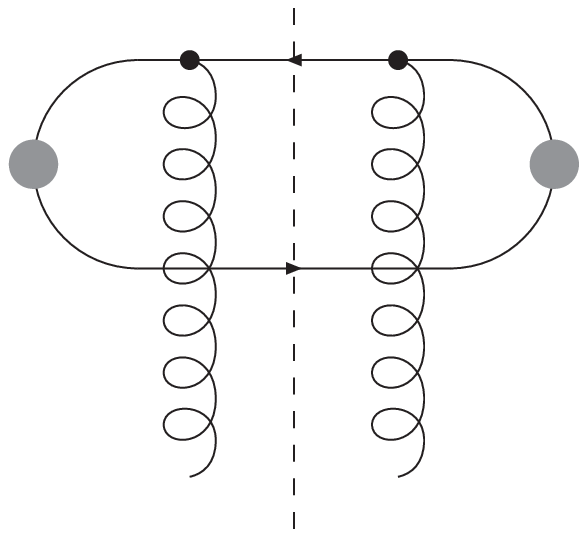}\\
    F4
  \end{minipage}
  \caption{The fermion diagrams for the impact factors.}
  \label{fig:IFFermions}
\end{figure}

\begin{figure}[htbp]
  \centering
  \begin{minipage}{2.5cm}
  \centering
    \includegraphics[width=2.5cm]{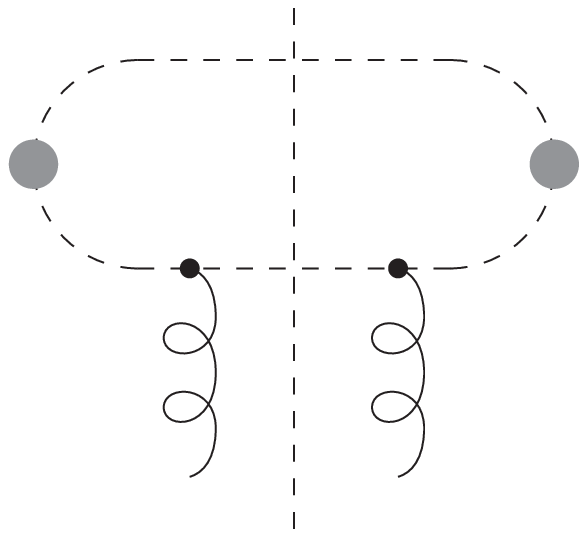}\\
    S1
  \end{minipage}\qquad
  \begin{minipage}{2.5cm}
  \centering
    \includegraphics[width=2.5cm]{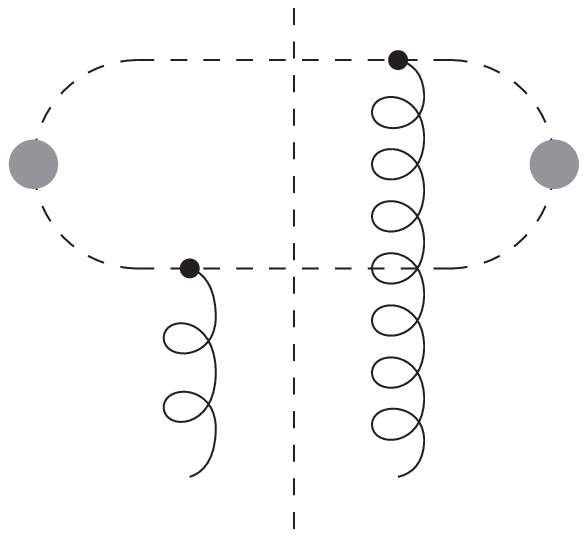}\\
    S2
  \end{minipage}\qquad
  \begin{minipage}{2.5cm}
  \centering
    \includegraphics[width=2.5cm]{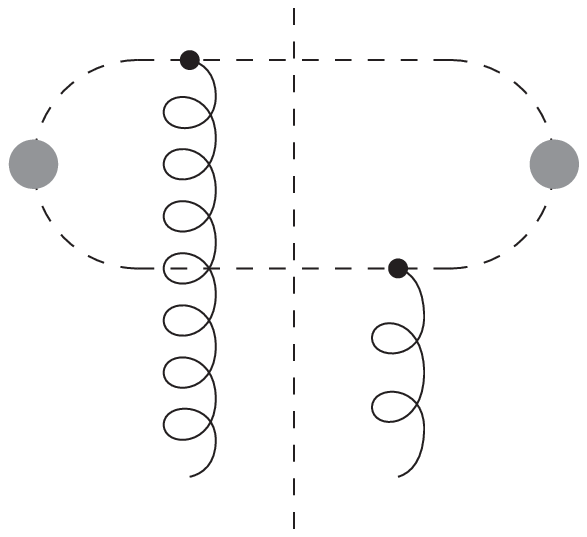}\\
    S3
  \end{minipage}\qquad
  \begin{minipage}{2.5cm}
  \centering
    \includegraphics[width=2.5cm]{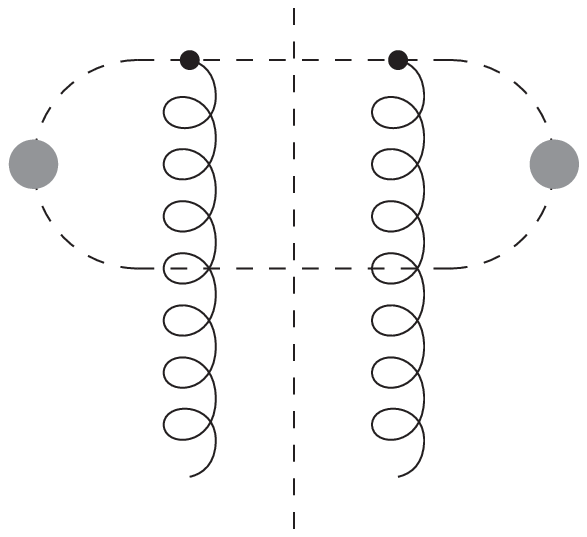}\\
    S4
  \end{minipage}\\[10pt]
  \begin{minipage}{2.5cm}
  \centering
    \includegraphics[width=2.5cm]{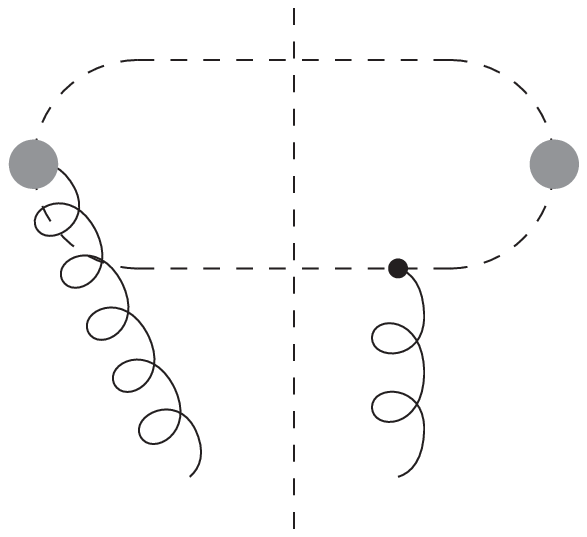}\\
    S5
  \end{minipage}\qquad
  \begin{minipage}{2.5cm}
  \centering
    \includegraphics[width=2.5cm]{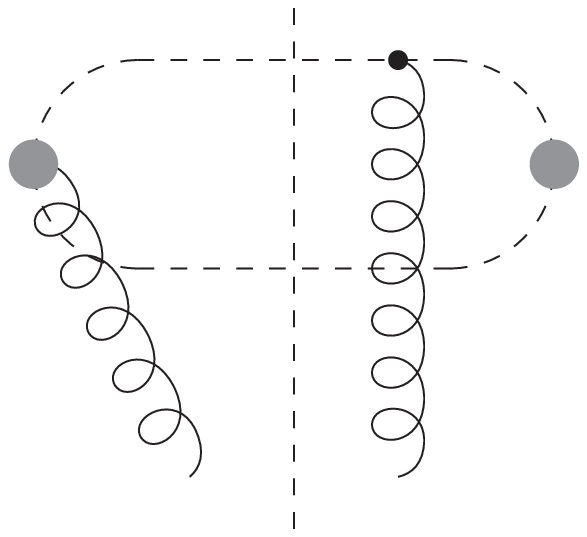}\\
    S6
  \end{minipage}\qquad
  \begin{minipage}{2.5cm}
  \centering
    \includegraphics[width=2.5cm]{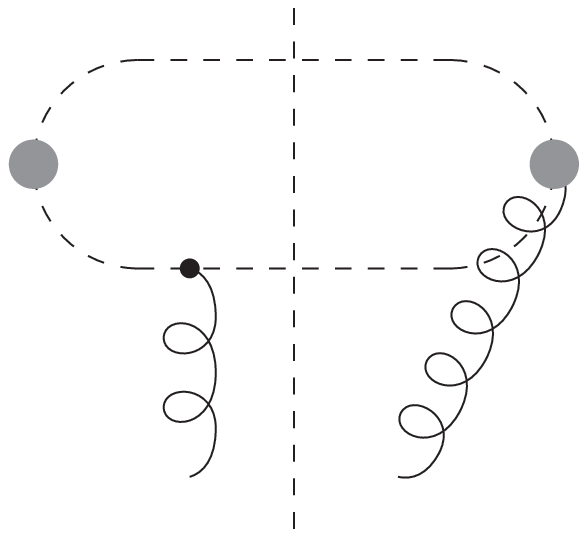}\\
    S7
  \end{minipage}\qquad
  \begin{minipage}{2.5cm}
  \centering
    \includegraphics[width=2.5cm]{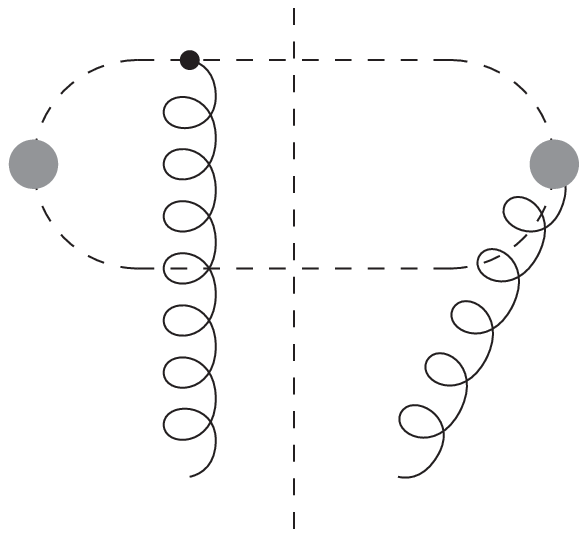}\\
    S8
  \end{minipage}\\[10pt]
  \begin{minipage}{2.5cm}
  \centering
    \includegraphics[width=2.5cm]{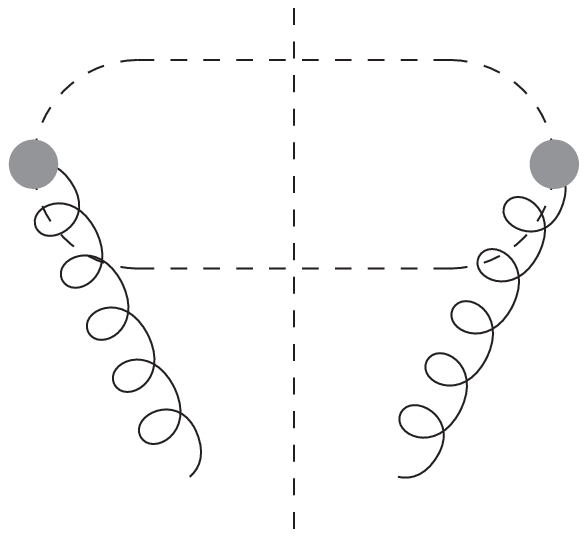}\\
    S9
  \end{minipage}
  \caption{The scalar diagrams for the impact factors.}
  \label{fig:IFScalars}
\end{figure}

It is convenient to introduce the notation
\begin{equation}
  \label{eq:PhiDecompDiagrams}
  \Phi^{\lambda \lambda' a a'} = N_c \alpha_s \delta^{a a'}
  \int_0^1d\alpha_l \int \frac{d^2 \bl}{(2\pi)^2}
  {\textstyle\sum_i} \phi^{\lambda \lambda'}_i(\alpha_l, \bl, \bq) \, ,
\end{equation}
where $\alpha_l$ is the longitudinal component of the (scalar or fermion)
loop integral along the incoming momentum $p_A$.
The term $\phi^{\lambda \lambda'}_i$ has to be computed from the diagram
$i$ in Figs. \ref{fig:IFFermions} and \ref{fig:IFScalars}.
The factor $N_c$ is present because both scalars and fermions 
belong to the adjoint representation of the gauge group, so they all give
\begin{equation}
  \label{eq:gaugeGroupFactor}
  f^{a c_1 c_2} f^{b c_2 c_1} = -N_c \delta^{a b} \, .
\end{equation}
An overall factor $1/2$ arises from the cutting rules,
$2 i \Im(A) = \Sigma \slashed{A}$.

The computation of the fermionic component (see Fig.\ \ref{fig:IFFermions}) is very similar to the
QCD case.
The first difference is due to the fact that in $\cN=4$
there are 4 Weyl fermions instead of $n_f$ Dirac ones.
The counting of the number of fields weighted by the right $R$-charge is
performed by the trace over the two generators of the \SUR~ group, 
\begin{equation}
  \label{eq:TATAfund}
  \Tr_4 \big( T^A T^A \big) = \frac{1}{2} \, ,
\end{equation}
(there is no sum over $A$ here),
taken in the appropriate representation (fundamental for the fermions and  vector representation for the 
scalars).

The chiral nature of the fields
introduces additional terms due to a Levi-Civita tensor arising from
spinor traces containing a chiral projector.
All these terms cancel in the sum of the four diagrams $F1-4$.
The complete list of the $\phi_i$ is given in the
appendix \ref{app:varphiList}.

The diagrams needed for the computation of the scalar component
of the impact factor are depicted in Fig.\ \ref{fig:IFScalars}.

The trace over the \SUR~ indices now gives 
\begin{equation}
  \label{eq:TATAvect}
  \Tr_6 \big( T^A T^A \big) = 1 \, .
\end{equation}
Since the scalars crossing the cut are identical particles there is 
a symmetry factor $1/2$.

At finite energies, all the diagrams $S1-9$ in Fig.\ \ref{fig:IFScalars} are needed
in order to satisfy the Ward identities. At high
energies, however, it turns out that the diagrams $S5-9$ are suppressed\footnote{
For simplicity, we discuss only the case of longitudinal polarization.}.
As an example, let us consider the diagram $S5$. The gluon polarization tensor is contracted 
with the polarization vector of the incoming current, which is proportional to $p_2$. 
Then, out of the three parts of the gluon polarization tensor 
(cf. \ref{eq:gluonPolLeading}: the index $\rho$ 
belongs to the upper end of the gluon, $\sigma$ to the lower end):
\begin{equation}
  \label{eq:gluonPolTens}
  \frac{2p_{2 \rho}p_{1\sigma}}{s} +
  \frac{2p_{1 \rho}p_{2\sigma}}{s} +
  g_{\perp \rho \sigma}\, ,
\end{equation}
only the second term survives because 
$p_2^2 = p_2^\mu g_{\perp \mu \nu}=0$.
Note that in the 'normal' fermionic case it is the \emph{first} term that gives the leading behavior
in the Regge asymptotic. One sees indeed that the contraction 
of the loop integral numerator with $p_1$ provides one power of $s$ less 
than the leading terms of diagrams $S1-4$. An analogous discussion  
applies to the contraction of $p_2$ with the loop below, 
and again there is a suppression
of a power of $s$. Eventually one sees that a diagram involving $S5$
is $1/s^2$ suppressed with respect to the leading term.
The same argument applies to the other diagrams $S6-8$, while for
$S9$ the suppression is even stronger, $1/s^4$, because the same effect
takes place for both gluons. We are thus left with the diagrams $S1-4$,
which, at high energies, give the full scalar component of the impact factor.
The computation mimics closely the fermionic one, and  details 
can be found in appendix \ref{app:varphiList}.

\subsection{The full impact factors}
Collecting together all the terms one obtains the full impact factors,
\begin{subequations}
  \label{eq:ImpFacN4}
  \begin{eqnarray}
    \label{eq:IFLL}
    \Phi_A^{L L' a a'}
 &=&\delta^{a a'} \frac{N_c \alpha_s}{2} Q_A Q_{A'}
    \!\!\int_0^1 \!\!d\alpha \!\!\int \!\!\frac{d^2\bl}{(2\pi)^2}
   \alpha (1-\alpha)\nonumber\\
    &&\times \Big( \frac{1}{D_{1}} - \frac{1}{D_{2}} \Big)
    \Big( \frac{1}{D'_{1}} - \frac{1}{D'_{2}} \Big) ,\\[5pt]
    \label{eq:IFLT}
    \Phi_A^{L h' a a'}
 &=&0 \, ,\\[5pt]
    \label{eq:IFTT}
    \Phi_A^{h h' a a'}
 &=&\delta^{a a'} \delta^{h h'} \frac{N_c \alpha_s}{2}
    \!\!\int_0^1 \!\!d\alpha \!\!\int \!\!\frac{d^2\bl}{(2\pi)^2}\nonumber\\
    &&\times\Big( \frac{\bN_1}{D_{1}} - \frac{\bN_2}{D_{2}} \Big) \cdot
    \Big( \frac{\bN'_{1}}{D'_{1}} - \frac{\bN'_2}{D'_{2}} \Big) \, ,
  \end{eqnarray}
\end{subequations}
with $D_i$ and $\bN_i$ defined by

\begin{subequations}
\begin{minipage}{7cm}
  \begin{eqnarray}
    \bN_{1} &=& \bl \, , \nonumber\\
    \bN'_{1} &=& \bl-(1-\alpha)\bq \, , \nonumber\\
    D_{1} &=& \bN_1^2 + \alpha (1-\alpha) Q_A^2 \, , \nonumber\\
    D'_{1} &=& \bN'^2_1 + \alpha (1-\alpha) Q_{A'}^2 \, , \nonumber
  \end{eqnarray}
  \begin{eqnarray}
    \label{eq:n2d2}
    \bN_{2} &=& \bl-\bk \, , \nonumber\\
    \bN'_{2} &=& \bl-\bk+\alpha\bq \, , \nonumber\\
    D_{2} &=& \bN_2^2 + \alpha (1-\alpha) Q_A^2 \, , \nonumber\\
    D'_{2} &=& \bN'^2_2 + \alpha (1-\alpha) Q_{A'}^2 \, . \nonumber
  \end{eqnarray}
\end{minipage}\\
\end{subequations}

Comparing (\ref{eq:ImpFacN4}a-c) with the QCD result of \cite{Bartels:2003yj}
one observes a striking difference: in contrast to the QCD results where 
helicity conservation holds only in the forward direction, at $t=0$, 
now for arbitrary $t=-\bq^2$ all the off-diagonal terms in the 
polarization indices vanish, as the result of cancellations between the 
scalar and fermion loops, 
$\Phi^{\lambda \lambda'} \propto \delta^{\lambda \lambda'} $. 
This is a consequence of supersymmetry.

Higher order diagrams with gluon exchange, in the LL approximation,  
lead to the QCD BFKL Pomeron described in Sec. \ref{sec:LL}. 
This coincidence, at high energies, of nonsupersymmetric Yang-Mills 
theory and the supersymmetric extension is an artifact of the leading 
logarithmic approximation, which only depends upon the $\mbox{spin-1}$ gauge 
bosons, and not on scalars or fermions. The only place where, in LL, these 
superpartners appear are the impact factors given now by
(\ref{eq:ImpFacN4}a-c). We have therefore completed our  
leading logarithmic analysis by proving that the correlation function 
\eqref{eq:4pointfunc} satisfies Regge factorization, and we have computed
those buildings blocks which are sensitive to the supersymmetric 
extension of QCD.  

\section{Outlook}
The AdS/CFT correspondence \cite{Maldacena:1997re-Witten:1998qj-Gubser:1998bc}
conjectures that $\cN=4$ SYM theory is equivalent to Type IIB superstring theory
on $AdS_5 \times S^5$.
The connection between these apparently different theories is a
weak-strong duality: it connects the weak coupling limit of one side
with the strong coupling limit on the other side.
This opens up the possibility to study aspects of the
gauge theory at strong coupling, where traditional tools are unapplicable.
In particular, we can address the computation of the $R$-currents correlation
function \eqref{eq:4pointfunc} in the large $N_c$ and large 't Hooft
coupling $\lambda = g_{YM}^2 N_c$ limit.
In this limit the relevant string theory is described by the $S^5$
compactification of type IIB supergravity in ten dimensions.
This reduction gives rise to $\cN=8$, $D=5$ supergravity, with $SO(6)$
Yang-Mills gauge group \cite{Schwarz:1983qr,Pernici:1984xx,Gunaydin:1984fk,
Kim:1985ez,Gunaydin:1984qu}.

The complete detailed reduction is a problem of great complexity.
Fortunately, there exist consistent truncations of the full theory which
are much simpler than the full theory. In \cite{Cvetic:1999xp} it was
shown that there is a very simple truncation which contains
only a $U(1)$ gauge field and the graviton.
Its action reads
\begin{eqnarray}
  \label{eq:U1truncation}
  &&e^{-1}\cL_5 =\\
&& - \frac{1}{2 \kappa_5^2} \Big(
  R + 12 g^2 -\frac{1}{4} F^2 + \frac{1}{12 \sqrt{3}}
  \epsilon^{\mu\nu\rho\sigma\lambda}
  F_{\mu\nu} F_{\rho\sigma} A_\lambda\nonumber
  \Big) \, .
\end{eqnarray}

According to the AdS/CFT dictionary, each gauge invariant operator
in the gauge theory corresponds to some bulk field in the supergravity theory.
The generating functional for the connected correlation functions
$W[\zeta]$ of the gauge theory, $\zeta$ being the source for some
operator $\cO$, is identified with the on-shell action $S_{\textrm{on-shell}}$
of the gravity theory, with the boundary conditions $\varphi_{(0)}$ for the
bulk field $\varphi$ dual to $\cO$ playing the role of its source $\zeta$:
\begin{equation}
  \label{eq:WeqS}
  W[\varphi_{(0)}] = -S_{\textrm{on-shell}}[\varphi_{(0)}] \, .
\end{equation}

The fields dual to the $R$-currents of $\cN=4$ SYM are the gauge fields
of the supergravity theory. The truncation \eqref{eq:U1truncation}
contains only one of the 15 gauge fields of the full theory,
in the same way as our computation in this paper concerns only 
one $R$-current of the U(1) subgroup out of the 15 associated with the \SUR~ group.
The action \eqref{eq:U1truncation} is therefore sufficient for
the purpose of computing the strong coupling version of our
result (\ref{eq:ImpFacN4}a-c).

\begin{figure}[htbp]
  \centering
    \includegraphics[width=3cm]{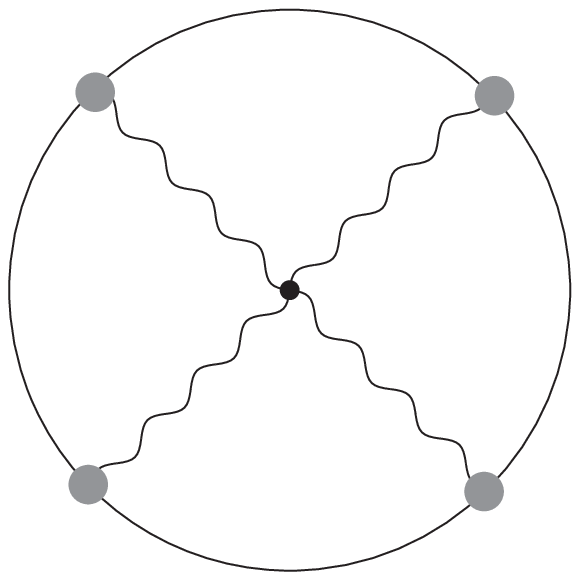}
    \includegraphics[width=3cm]{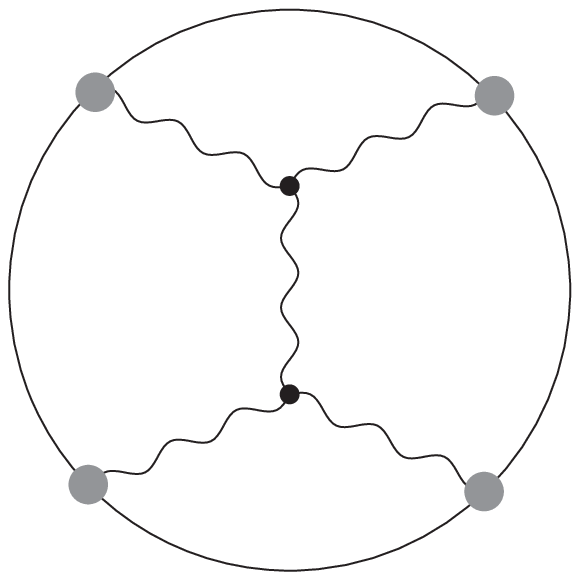}
    \includegraphics[width=3cm]{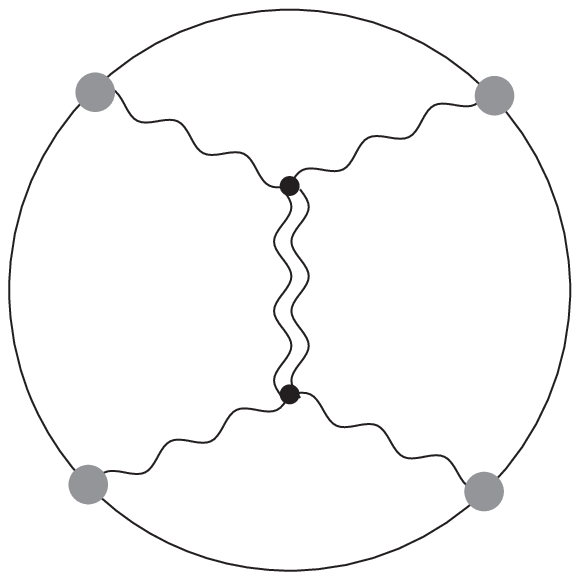}
  \caption{Witten diagrams for the computation of the $R$-current four
    point function at strong coupling in the truncated theory.
    The double wavy line in the third graph represents the graviton
    exchange.}
  \label{fig:WD}
\end{figure}

The supergravity computation requires the evaluation of the
Witten diagrams corresponding to some sources for the gauge field
$A_\mu$ on the boundary of $AdS_5$. Diagrams inferred from the
action \eqref{eq:U1truncation} are depicted in Fig.\ \ref{fig:WD}.
Such computation involves the boundary-to-bulk gauge boson propagator
and the bulk-to-bulk propagators for both the gauge field and
the graviton.
They are well known in the coordinate space \cite{D'Hoker:1999jc},
and have been extensively used in the past, in order to compute
various correlation functions (see for example \cite{D'Hoker:1999pj}
and references therein). Nevertheless the computation of a four point
correlation function of $R$-currents is still missing in the literature.
We intend to address this computation in the future,
not in its full generality but in the Regge limit, where we expect
some simplifications to take place.

Returning to the gauge theory side, our analysis of the supersymmetric 
$R$-current impact factors lays the ground for addressing another 
aspect of the duality conjecture. Several years ago it has been shown that 
the BKP evolution equations of $t$-channel states consisting of 
$n$ Reggeized gluons, in the limit of large $N_c$, are integrable 
\cite{Lipatov:1993yb,LiFK}.
On the gauge theory side, the four gluon state appears in the high energy 
limit of the six point correlation function of $R$-currents
(in QCD, the analogous process would be the scattering of a virtual 
photon on two heavy onium states). A study of this correlation function,
both on the weak coupling and on the strong coupling side, therefore 
will allow us to trace the role of this remarkable feature of the 
Regge limit.\\ \\ 

\begin{acknowledgments} We thank L. N. Lipatov for very helpful discussions. A.-M. M. is supported by the Graduiertenkolleg 
``Zuk\"unftige Entwicklungen in der Teilchenphysik''. \mbox{M. S.} has been supported by the Sonderforschungsbereich ``Teilchen, Strings und Fr\"uhes Universum''.
\end{acknowledgments}


\begin{appendix}

\section{}

\subsection{Sudakov decomposition and polarization vectors}
\label{app:sudakovAndPol}

We will work in the frame where $p_A$ and $p_A'$ ($p_B$ and $p_B'$)
have a big $+$ ($-$) component, with $p_{\pm}=p_0 \pm p_3$.
In this basis the nonvanishing metric coefficients are \mbox{$g_{+-}=g_{-+}=2$}
and $g_{11}=g_{22}=1$, while the Levi-Civita tensor is fully antisymmetric
with $\epsilon^{+-12}=-1/2$.
In the c.m. we have\footnote{Only the leading term in $s$ is kept
for each component.}
\begin{equation}
  \label{eq:momenta}
    p_A = \Big(\sqrt{s}, -\frac{Q_A^2}{\sqrt{s}}, 0 \Big) \, ,\qquad
    p_B = \Big(-\frac{Q_B^2}{\sqrt{s}}, \sqrt{s}, 0 \Big) \, .
\end{equation}
We use the Sudakov representation for momenta. Let us define the two
lightlike (up to $\cO(1/s)$ terms) vectors $p_1 = p_A + x_A p_B$,
$p_2 = p_B + x_B p_A$ where \mbox{$x_{A,B} = Q_{A,B}^2/2 p_A \cdot p_B$}.
Then an arbitrary four-momentum $k$ will be decomposed into its
projections along $p_{1,2}$ and a transverse component:
\begin{equation}
  \label{eq:sudakovDec}
  k=\alpha_k p_1 + \beta_k p_2 + k_{\perp}, \qquad\qquad
  \Big\{\begin{array}{ccc}
    \alpha_k &=& 2 p_2 \cdot k / s \\
    \beta_k &=& 2 p_1 \cdot k / s.
  \end{array}
\end{equation}
The Jacobian is
\begin{equation}
  \label{eq:d4k}
  d^4k = s/2 d\alpha_k d\beta_k d^2\bk \, .
\end{equation}
The mass-shell conditions for the outgoing momenta fix the longitudinal
components of the momentum transferred $q$ to
\begin{equation}
  \label{eq:AlphaqBetaq}
  \alpha_q = -\frac{Q_{B'}^2-Q_B^2+q_{\perp}^2}{s} \, , \qquad\qquad
  \beta_q = \frac{Q_{A'}^2-Q_A^2+q_{\perp}^2}{s} \, ,
\end{equation}
and the external momenta expressed in Sudakov representation are
\begin{equation}
  \label{eq:momentaSudakov}
  \begin{array}{ll}
    p_A = p_1 - \frac{Q_{A}^2}{s} p_2 \, ,&\qquad
    p_A' = p_1 - \frac{Q_{A'}^2 + q_{\perp}^2}{s} p_2 - q_{\perp} \, ,\\
    p_B = -\frac{Q_B^2}{s} p_1 + p_2 \, ,&\qquad
    p_B' = -\frac{Q_{B'}^2 + q_{\perp}^2}{s} p_1 + p_2 + q_{\perp} \, .
  \end{array}
\end{equation}

We now compute the polarization vectors of the external photons. 
Virtual photons have 3 degrees of freedom,
one longitudinal (L) and two transverse ($\pm$) polarizations. The absence of a second longitudinal polarization
translates for the amplitude \eqref{eq:4pointfuncQCD} into the constraint
\eqref{eq:WardId4} provided by gauge invariance.
Because of this constraint, the three polarization vectors $\epsilon^{L,\pm}_\mu(p)$
represent, for an arbitrary choice of the momentum $p$, 
a complete basis of the space where the current $j_\mu(p)$
belongs to
\begin{equation}
  \label{eq:jspan}
  j_\mu(p) \in \textrm{Span}\big\{ \epsilon^{(L,\pm)}_\mu(p) \big\} \, ,
\end{equation}
They can be chosen to be orthonormal,
\begin{equation}
  \label{eq:epsilonOrth}
  \epsilon^{(i)}_\mu(p) \epsilon^{(j) \mu}(p)^* = - \delta^{ij} \, ,
\end{equation}
and to satisfy the completeness relation
\begin{equation}
  \label{eq:polComp}
  g^{\mu \nu} - \frac{p^\mu p^\nu}{p^2} =
  \sum_{i=L,\pm} \epsilon^{(i)\mu}(p) \epsilon^{(i)\nu}(p)^* \, .
\end{equation}
We choose $\epsilon^L_\mu(p)$ such that its three-dimensional
part is proportional to the three-momentum $\vec{p}$
(longitudinal polarization). The two other vectors (helicity $\pm$)
are chosen to be transverse. In the Sudakov representation
(keeping only the leading term in $s$ for each component) we get
\begin{eqnarray}
  \label{eq:polVects}
    &&\epsilon^{(L)} (p) =\nonumber\\
    &&\quad\frac{i}{Q} \Big[
    \Big( \alpha + \frac{2 Q^2}{s (\alpha + \beta)^2} \beta \Big) p_1 +
    \Big( \beta + \frac{2 Q^2}{s (\alpha + \beta)^2} \alpha \Big) p_2\nonumber\\
    &&\quad+ \Big( 1- \frac{2 Q^2}{s (\alpha + \beta)^2} \Big) p_\perp \Big] \, ,\\
    &&\epsilon^{(h)} (p) = \epsilon^{(h)}_\perp +
    \frac{2 \epsilon^{(h)}_\perp \cdot p}{s(\alpha-\beta)}
    \Big( p_1 - p_2 + \frac{p_\perp}{\alpha - \beta} \Big) \, ,
\end{eqnarray}
where we have defined
\begin{eqnarray}
  \label{eq:epsilonhperp}
  \epsilon^{(\pm)}_\perp = \frac{1}{\sqrt{2}} (0,1,\pm i,0) \, .
\end{eqnarray}
The explicit expressions for the case $p = p_A, p_{A'}, p_B, p_{B'}$
can be easily worked out from \eqref{eq:momentaSudakov}:
\begin{eqnarray}
  \label{eq:polVecExplicitLA}
  \epsilon^{(L)}(p_A) &=&
  \frac{i}{Q_A} \Big( p_1 + \frac{Q_A^2}{s} p_2 \Big) \, , \\
  \label{eq:polVecExplicitLB}
  \epsilon^{(L)}(p_B) &=&
  \frac{i}{Q_B} \Big( \frac{Q_B^2}{s} p_1 + p_2 \Big) \, , \\
  \label{eq:polVecExplicitLAp}
  \epsilon^{(L)}(p_{A'}) &=&
  \frac{i}{Q_{A'}} \Big( p_1 + \frac{Q_{A'}^2 - q_{\perp}^2}{s} p_2
  - q_\perp \Big) , \\
  \label{eq:polVecExplicitLBp}
 \epsilon^{(L)}(p_{B'}) &=&
  \frac{i}{Q_{B'}} \Big( \frac{Q_{B'}^2 - q_{\perp}^2}{s} p_1 + p_2
  + q_\perp \Big)  , \\
  \label{eq:polVecExplicithAB}
  \epsilon^{(h)}(p_{A,B}) &=& \epsilon^{(h)}_\perp \, , \\
  \label{eq:polVecExplicithApBp}
  \epsilon^{(h)}(p_{A',B'}) &=& \epsilon^{(h)}_\perp
  - \frac{2 \epsilon^{(h)}_\perp \cdot q}{s} (p_1 - p_2 - q_\perp) \, .
\end{eqnarray}
Because of the Ward identities \eqref{eq:WardId4} and \eqref{eq:WardId6}
one is allowed to shift the polarization vectors $\epsilon^{(i)}_\mu(p)$
by a four vector proportional to $p$ itself. It is convenient 
to simplify the polarization vectors as follows:
\begin{eqnarray}
  \label{eq:polVecReducedLAAp}
  \epsilon^{(L)}(p_{A,A'}) &=& \frac{2 Q_{A,A'}}{s} p_2 \, , \\
  \label{eq:polVecReducedLBBp}
  \epsilon^{(L)}(p_{B,B'}) &=& \frac{2 Q_{B,B'}}{s} p_1 \, , \\
  \label{eq:polVecReducedhAB}
  \epsilon^{(h)}(p_{A,B}) &=& \epsilon^{(h)}_\perp \, , \\
  \label{eq:polVecReducedhAp}
  \epsilon^{(h)}(p_{A'}) &=& \epsilon^{(h)}_\perp
  + \frac{2 \epsilon^{(h)}_\perp \cdot q}{s}  p_2 \, , \\
  \label{eq:polVecReducedhBp}
  \epsilon^{(h)}(p_{B'}) &=& \epsilon^{(h)}_\perp
  - \frac{2 \epsilon^{(h)}_\perp \cdot q}{s}  p_1 \, .
\end{eqnarray}

\subsection{Complete list of the $\phi^{ii'}_{S,Fl}$}
\label{app:varphiList}

In this appendix we give, for all possible polarizations 
$\lambda,\lambda'=L,\pm$, the full list of the functions
$\phi_i^{\lambda \lambda'}$ for the eight diagrams of
Figs. \ref{fig:IFFermions} ($i=F1-4$) and \ref{fig:IFScalars} ($i=S1-4$).
We will make use of the definitions for $D_i$ and $N_i$,
given after \eqref{eq:ImpFacN4}.

\subsubsection*{Longitudinal-Longitudinal:}
  \begin{eqnarray}
    \label{eq:phisListLLF}
    \phi_{F1}^{LL'} &=&
    2 Q_A Q_{A'} \frac{\alpha^2 (1-\alpha)^2}{D_1 D'_1} \, , \nonumber\\
    \phi_{F2}^{LL'} &=&
    -2 Q_A Q_{A'} \frac{\alpha^2 (1-\alpha)^2}{D_1 D'_2} \, , \nonumber\\
    \phi_{F3}^{LL'} &=&
    -2 Q_A Q_{A'} \frac{\alpha^2 (1-\alpha)^2}{D_2 D'_1} \, , \nonumber\\
    \phi_{F4}^{LL'} &=&
    2 Q_A Q_{A'} \frac{\alpha^2 (1-\alpha)^2}{D_2 D'_2} \, , 
  \end{eqnarray}
  \begin{eqnarray}
    \label{eq:phisListLLS}
    \phi_{S1}^{LL'} &=&
    2 Q_A Q_{A'} \frac{\alpha (1-\alpha) (1/2-\alpha)^2}{D_1 D'_1}
    \, , \nonumber\\
    \phi_{S2}^{LL'} &=&
    -2 Q_A Q_{A'} \frac{\alpha (1-\alpha) (1/2-\alpha)^2}{D_1 D'_2}
    \, , \nonumber\\
    \phi_{S3}^{LL'} &=&
    -2 Q_A Q_{A'} \frac{\alpha (1-\alpha) (1/2-\alpha)^2}{D_2 D'_1}
    \, , \nonumber\\
    \phi_{S4}^{LL'} &=&
    2 Q_A Q_{A'} \frac{\alpha (1-\alpha) (1/2-\alpha)^2}{D_2 D'_2}
    \, . 
  \end{eqnarray}

\subsubsection*{Longitudinal-Transverse:}
  \begin{eqnarray}
    \label{eq:phisListLTF}
    \phi_{F1}^{Lh'} &=&
    Q_A \frac{\alpha (1-\alpha) (1-2\alpha-h'i)}{D_1 D'_1}
    \bN'_1 \cdot \beps^{(h')*} \, , \nonumber\\
    \phi_{F2}^{Lh'} &=&
    -Q_A \frac{\alpha (1-\alpha) (1-2\alpha-h'i)}{D_1 D'_2}
    \bN'_2 \cdot \beps^{(h')*} \, , \nonumber\\
    \phi_{F3}^{Lh'} &=&
    -Q_A \frac{\alpha (1-\alpha) (1-2\alpha-h'i)}{D_2 D'_1}
    \bN'_1 \cdot \beps^{(h')*} \, , \nonumber\\
    \phi_{F4}^{Lh'} &=&
    Q_A \frac{\alpha (1-\alpha) (1-2\alpha-h'i)}{D_2 D'_2}
    \bN'_1\cdot \beps^{(h')*} \, ,
  \end{eqnarray}
  \begin{eqnarray}
    \label{eq:phisListLTS}
    \phi_{S1}^{Lh'} &=&
    -Q_A \frac{\alpha (1-\alpha) (1-2\alpha)}{D_1 D'_1}
    \bN'_1 \cdot \beps^{(h')*} \, , \nonumber\\
    \phi_{S2}^{Lh'} &=&
    Q_A \frac{\alpha (1-\alpha) (1-2\alpha)}{D_1 D'_2}
    \bN'_2 \cdot \beps^{(h')*} \, , \nonumber\\
    \phi_{S3}^{Lh'} &=&
    Q_A \frac{\alpha (1-\alpha) (1-2\alpha)}{D_2 D'_1}
    \bN'_1 \cdot \beps^{(h')*} \, , \nonumber\\
    \phi_{S4}^{Lh'} &=&
    -Q_A \frac{\alpha (1-\alpha) (1-2\alpha)}{D_2 D'_2}
    \bN'_2 \cdot \beps^{(h')*} \, . 
  \end{eqnarray}

Since under  the
transformations $\bl \rightarrow -\bl+\bk$ and $\alpha\rightarrow1-\alpha$, 
$N_2 \rightarrow -N_1$ and $N'_2 \rightarrow -N'_1$ 
in the integrand in \eqref{eq:PhiDecompDiagrams},
the terms proportional to the helicity $h'$ in the fermionic parts 
cancel between $\phi_{F1}^{Lh'},\phi_{F4}^{Lh'}$  
and $\phi_{F2}^{Lh'},\phi_{F3}^{Lh'}$.
The remaining fermionic pieces cancel completely against the 
corresponding scalar terms.

\subsubsection*{Transverse-Longitudinal:}
  \begin{eqnarray}
    \label{eq:phisListTLF}
    \phi_{F1}^{hL} &=&
    Q_{A'} \frac{\alpha (1-\alpha) (1-2\alpha-hi)}{D_1 D'_1}
    \beps^{(h)} \cdot \bN_1 \, , \nonumber\\
    \phi_{F2}^{hL} &=&
    -Q_{A'} \frac{\alpha (1-\alpha) (1-2\alpha-hi)}{D_1 D'_2}
    \beps^{(h)} \cdot \bN_1 \, , \nonumber\\
    \phi_{F3}^{hL} &=&
    -Q_{A'} \frac{\alpha (1-\alpha) (1-2\alpha-hi)}{D_2 D'_1}
    \beps^{(h)} \cdot \bN_2 \, , \nonumber\\
    \phi_{F4}^{hL} &=&
    Q_{A'} \frac{\alpha (1-\alpha) (1-2\alpha-hi)}{D_2 D'_2}
    \beps^{(h)} \cdot \bN_2 \, ,
  \end{eqnarray}
  \begin{eqnarray}
    \label{eq:phisListTLS}
    \phi_{S1}^{hL} &=&
    -Q_{A'} \frac{\alpha (1-\alpha) (1-2\alpha)}{D_1 D'_1}
    \beps^{(h)} \cdot \bN_1 \, , \nonumber\\
    \phi_{S2}^{hL} &=&
    Q_{A'} \frac{\alpha (1-\alpha) (1-2\alpha)}{D_1 D'_2}
    \beps^{(h)} \cdot \bN_1 \, , \nonumber\\
    \phi_{S3}^{hL} &=&
    Q_{A'} \frac{\alpha (1-\alpha) (1-2\alpha)}{D_2 D'_1}
    \beps^{(h)} \cdot \bN_2 \, , \nonumber\\
    \phi_{S4}^{hL} &=&
    -Q_{A'} \frac{\alpha (1-\alpha) (1-2\alpha)}{D_2 D'_2}
    \beps^{(h)} \cdot \bN_2 \, .
  \end{eqnarray}
Here we have the same cancellations as in the longitudinal-trans\-verse
case.
\subsubsection*{Transverse-Transverse:}
\begin{eqnarray}
  \label{eq:phisListTTF}
  \phi_{F1}^{hh'} &=&
  \frac{1}{2D_1 D'_1} \Big[ (1-hi\alpha) ~ \beps^{(h)} \cdot \beps^{(h')*}~
  \bN_1 \cdot \bN'_1 \nonumber\\
 && + \Big(- 4 \alpha (1-\alpha) + i(h - h') (1-\alpha) \Big) ~\nonumber\\
 &&\qquad\qquad\qquad\qquad\qquad\times \beps^{(h)} \cdot \bN_1 ~
  \bN'_1 \cdot \beps^{(h')*} \nonumber\\
 &&- (1 - h i \alpha) \Big(
  \beps^{(h)} \cdot \bl ~ (1-\alpha)\bq \cdot \beps^{(h')*}\nonumber\\
&&\qquad\qquad\qquad\qquad - \beps^{(h)} \cdot (1-\alpha)\bq ~ \bl \cdot \beps^{(h')*}
  \Big)
  \Big] \, ,\nonumber \\
  \phi_{F2}^{hh'} &=&
  -\frac{1}{2D_1 D'_2} \Big[ \beps^{(h)} \cdot \beps^{(h')*}~
  \bN_1 \cdot \bN'_2 \nonumber\\
 && + \Big( - 4 \alpha(1-\alpha) + i(h-h') (1-2\alpha)\Big) ~\nonumber\\
 &&\qquad\qquad\qquad\qquad\qquad \times \beps^{(h)} \cdot \bN_1 ~
  \bN'_2 \cdot \beps^{(h')*} \nonumber\\
 &&  - \beps^{(h)} \cdot \bl ~ (\bk - \alpha \bq) \cdot \beps^{(h')*}\nonumber\\
&&\qquad\qquad\qquad\qquad+ \beps^{(h)} \cdot (\bk - \alpha \bq) ~ \bl \cdot \beps^{(h')*}
  \Big] \, , \nonumber\\
  \phi_{F3}^{hh'} &=&
  -\frac{1}{2D_2 D'_1} \Big[ \beps^{(h)} \cdot \beps^{(h')*}~
  \bN_2 \cdot \bN'_1 \nonumber\\
&&  + \Big( - 4 \alpha (1-\alpha) + i(h - h') (1 - 2 \alpha)\Big)\nonumber\\
&&\qquad\qquad\qquad\qquad\qquad\times  \beps^{(h)} \cdot \bN_2 ~
  \bN'_1 \cdot \beps^{(h')*}  \nonumber\\
&& +\beps^{(h)} \cdot (\bl - \bk) ~ (\bk-(1 - \alpha)\bq) \cdot \beps^{(h')*}\nonumber\\
&& \qquad\quad-\beps^{(h)} \cdot (\bk - (1-\alpha)\bq) ~ (\bl - \bk) \cdot \beps^{(h')*}
  \Big] \, , \nonumber\\
  \phi_{F4}^{hh'} &=&
  \frac{1}{2D_2 D'_2} \Big[ (1+hi(1-\alpha)) ~ \beps^{(h)} \cdot \beps^{(h')*}~
  \bN_2 \cdot \bN'_2 \nonumber\\
&&  + \Big( - 4 \alpha(1-\alpha) - i(h-h') (1-\alpha)\Big) ~\nonumber\\
 &&\qquad\qquad\qquad\qquad\qquad\times \beps^{(h)} \cdot \bN_2~
  \bN'_2 \cdot \beps^{(h')*} \nonumber\\
 && +(1+hi(1-\alpha)) \Big( \beps^{(h)} \cdot (\bl - \bk) ~
  \alpha\bq \cdot \beps^{(h')*}\nonumber\\
&& \qquad\qquad\quad- \beps^{(h)} \cdot \alpha\bq ~ (\bl - \bk) \cdot \beps^{(h')*} \Big)
  \Big] \, ,
\end{eqnarray}
\begin{eqnarray}
  \phi_{S1}^{hh'} &=&
  2 \frac{\alpha (1-\alpha)}{D_1 D'_1}
  \beps^{(h)} \cdot \bN_1 ~ \bN'_1 \cdot \beps^{(h')*}
  \, , \nonumber\\
  \phi_{S2}^{hh'} &=&
  -2 \frac{\alpha (1-\alpha)}{D_1 D'_2}
  \beps^{(h)} \cdot \bN_1 ~ \bN'_2 \cdot \beps^{(h')*}
  \, , \nonumber\\
  \phi_{S3}^{hh'} &=&
  -2 \frac{\alpha (1-\alpha)}{D_2 D'_1}
  \beps^{(h)} \cdot \bN_2 ~ \bN'_1 \cdot \beps^{(h')*}
  \, , \nonumber\\
  \phi_{S4}^{hh'} &=&
  2 \frac{\alpha (1-\alpha)}{D_2 D'_2}
  \beps^{(h)} \cdot \bN_2 ~ \bN'_2 \cdot \beps^{(h')*}
  \, . 
\end{eqnarray}

Here the cancellations are a bit more involved. 
In the fermionic sector of each $\phi$, 
the two terms in the last line cancel each other due to
the angular integration in the transverse momenta $\bl$. In order
to see this one combines the two denominators introducing
a Feynman parameter and then performs a shift in the $\bl$ integration.
The shift in the numerator cancels between the two terms, and what is left
depends upon the angle in the transverse plane only through the 
$\cos (\theta)$ in the scalar product with the polarization vectors 
in the numerator; therefore the $\theta$ integral vanishes.

From what is left, all the terms proportional to the helicities $h,h'$
cancel in the same way as they did in the previous case:
between $\phi_{F1}^{hh'},\phi_{F4}^{hh'}$  and
$\phi_{F2}^{hh'},\phi_{F3}^{hh'}$ after the change of variable
$\bl \rightarrow -\bl+\bk$ and $\alpha \rightarrow 1-\alpha$.
Moreover, the terms from the scalar sector cancel exactly with the
corresponding terms in the fermionic sector. Eventually only
a single term proportional to $\beps^{(h)} \cdot \beps^{(h')*}$
is left for each diagram.

\end{appendix}

\end{document}